\documentclass[%
 reprint,
 amsmath,amssymb,
 aps,
floatfix
]{revtex4-1}

\usepackage{graphicx}
\usepackage{dcolumn}
\usepackage{bm}
\usepackage[colorlinks=true]{hyperref}
\usepackage{siunitx}
\usepackage{subfigure}
\usepackage[T1]{fontenc}
\usepackage{float}

\usepackage{xcolor}
\usepackage{soul}
\usepackage{systeme}
\newcommand{\erfc}{\text{erfc}}

\begin{document}




\title{Molecular scale ion separation driven by surface roughness and ion size asymmetry: new analytical solutions for differential capacitance of EDL}
%
\author{Aleksey Khlyupin}
\email{khlyupin@phystech.edu}
\affiliation{Moscow Institute of Physics and Technology, Institutskiy Pereulok 9, Dolgoprudny, Moscow, Russia 141700}
\affiliation{Schmidt Institute of Physics of the Earth of Russian Academy of Sciences, Bolshaya Gruzinskaya 10, Moscow, Russia, 123242}

\author{Irina Nesterova}
\email{irina.nesterova@phystech.edu}
\affiliation{Moscow Institute of Physics and Technology, Institutskiy Pereulok 9, Dolgoprudny, Moscow, Russia 141700}
\affiliation{Schmidt Institute of Physics of the Earth of Russian Academy of Sciences, Bolshaya Gruzinskaya 10, Moscow, Russia, 123242}

\author{Kirill Gerke}
\email{kg@ifz.ru}
\affiliation{Schmidt Institute of Physics of the Earth of Russian Academy of Sciences, Bolshaya Gruzinskaya 10, Moscow, Russia, 123242}

\date{\today}

\begin{abstract}

Electrode surface roughness significantly impacts the structure of electric double layer on a molecular scale. We derive analytical solutions for differential capacitance (DC) of electric double layer near rough electrode surface, comparing them with a range of experimental and numerical studies. Two causes of ions separation are considered: ion size asymmetry and electrode surface roughness. The model has three scale parameters determining DC properties: the Debye length, difference of ion penetration depths, and surface roughness parameter. For the first time, DC profile with more than two peaks was obtained analytically due to account for ions reorientation effect. The model predicts DC curve transform from bell to camel and inverse induced by electrode surface roughness. The behavior of DC-potential relation with temperature and ion size ratio diverges depending on the type of ions separation. Additionally, we provide analytical solutions for zero charge potential and demonstrate roughness effect on its value. Based on the results, we give recommendations for properties required to design effective electric double layer capacitors.

\end{abstract}

\maketitle
\section{Introduction}

The extensive research intended on energy storage tools design is in great demand, as it will enable to operate electric power systems efficiently\cite{wallace2009nanoelectrodes, liu2010advanced, gur2018review, yang2018battery, zhang2018energy, gielen2019role}. One of the most prospective technologies for energy storage systems is the electric double layer capacitor (EDLC) with the involvement of room temperature ionic liquids (RTIL) \cite{galinski2006ionic, chen2009progress, pech2010ultrahigh, fedorov2014ionic, gonzalez2016review}. To establish high capacitance characteristics of EDLC, it is essential to understand and predict the behavior of the electric double layer near the electrode surface. Since numerical simulations are computationally expensive, it is required to develop analytical models representing the influence of electrode surface and ionic liquid (IL) physical properties on the differential capacitance of EDLC \cite{daikhin1996double, daikhin1998nonlinear, kornyshev2007double, wang2011accurate, chen2018temperature}.

Plurality of previous experimental, simulation, and theoretical studies demonstrated that the capacitance-potential curve for ionic liquids differs from classical electrochemical Goy\,--\,Chapman relation due to stronger ion interactions and direct contact of ions with electrode surface as there is no solvent phase. For example, Nanjundiah et.al \cite{nanjundiah1997differential} were the first, who obtained this inconsistency experimentally. Then, Kornyshev \cite{kornyshev2007double} proposed a theory that treats a finite volume of ions and provides bell and camel-like DC profiles, which was proved by molecular simulations \cite{fedorov2008ionic}. The nature of the camel shape profile was explained by ion dispersion forces \cite{trulsson2010differential} and by the presence of neutral parts in the ions structure \cite{fedorov2010double}. 

\begin{figure*}
    \centering
    \includegraphics[width=1\linewidth]{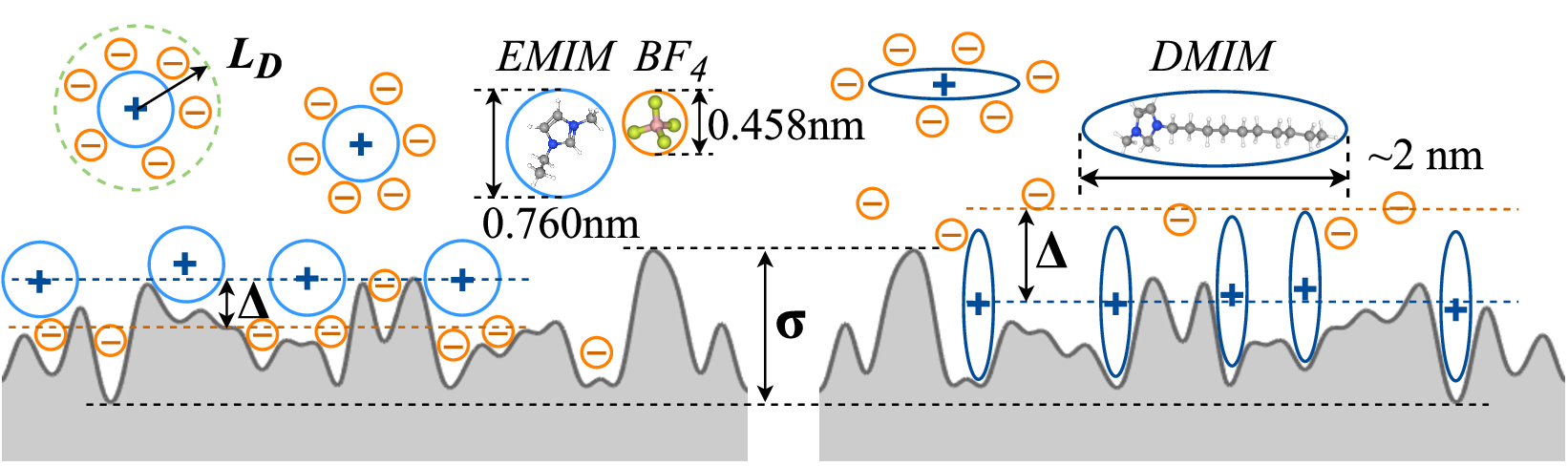}
    \caption{Schematic illustration of scale parameters that rule ion separation process: Debye length $L_D$, difference of ion penetration depth $\Delta$ and roughness parameter $\sigma$. On the left, homogeneous state of IL ions is presented, when smaller ions penetrate deeper than bigger one. On the right, we demonstrate inhomogeneous state, where long big cations become perpendicularly oriented to electrode surface and start to penetrate deeper than smaller ions. Here, 1-ethyl-3-methylimidazolium EMIM$^+$,  1-decyl-3-methyl imidazolium DMIM$^+$ and tetrafluoroborate BF$_4^-$ ions are considered.}
    \label{fig:TOC}
\end{figure*} 

However it is still challenging to describe all specificity of ion interactions resulting in DC of electric double layer (EDL) especially on surfaces with molecular scale roughness. Regarding the surface impact on EDL structure, Daikhin and coworkers derived the analytical expression for DC that accounts impact of electrode surface roughness \cite{daikhin1996double}. They first solved the Poisson-Boltzmann equation in the linear approximation of small potentials with a boundary condition defined on rough solid surface (macroscopic roughness). The obtained result is limited by small roughness, potential, and ion concentration. Later, authors extended this approach by solving the nonlinear Poisson-Boltzmann equation, so they overcame the low potential limitation \cite{daikhin1998nonlinear}. Though, in these works roughness is geometrical macro characteristic that improves capacitance due to the raise of surface area and electric field redistribution. It cannot describe hard repulsion interactions between ions and solid surface that dramatically influence EDL structure in the case of nanoscale roughness.

Recent molecular dynamic studies found that molecular scale roughness of electrode surface leads to significant change of interfacial layer structure, i.e modifies ions density distribution, changes ions orientation near electrode surface, and speeds up ion redistribution with potential alteration \cite{vatamanu2011influence, vatamanu2012molecular, xing2012nanopatterning, hu2013molecular, hu2014comparative, lu2017structure}. The revealed effects modify DC profile form and create in order higher DC variation with potential. However, it is still less clear, especially from theoretical point of view, how to treat molecular scale roughness, due to controversial results and inconsistency of experimental data \cite{vatamanu2011influence, hu2013molecular}. Therefore, a theory-based explanation of these phenomena is highly desired. More recently, a new theoretical model for ion density distribution was formulated in terms of statistical mechanics using mean-field approach and regarding solid surface as a correlated random process \cite{aslyamov2021electrolyte}, following the work, where this rough surface model was proposed \cite{khlyupin2017random}. The model treats electrode roughness on a molecular scale and represents DC enhancement by the increase of effective electric field and stronger ions separation. However, the model is too complicated that DC can be evaluated only numerically. Therefore, it is still important to develop an analytical models for DC that will account roughness on a molecular scale and will provide an easier analysis of capacitance characteristics.

This article provides new class of analytical solutions for DC of EDL on surface roughness accounted on a molecular scale. The model is based on 3 scale parameters, that rule EDL structure: the Debye length $L_D$, roughness parameter of electrode surface $\sigma$, and difference of ion penetration depths $\Delta$, which are shown in Fig.\ref{fig:TOC}. Depending on the relation of these parameters, we distinguish two mechanisms for ion separation. In the first one, called the ion size separation (ISS) model, ions separate due to large difference of ion penetration depths that occurs for ions with large difference of ion sizes and small electrode roughness. In the second model, called the roughness separation (RS) model, ion separation happens due to a large roughness of electrode surface that is much bigger than difference of ions penetration depths. Here, we solve nonlinear Poisson-Boltzmann for perturbation caused by surface roughness. Thus, we obtain analytical solutions for DC with the influence of repulsion interactions between rough electrode surface and IL ions within simple 1D models that account ion sizes and different types of ion separation. We use two perturbation methods due to different nature of the small parameter in both models.

The results of the presented models cover a wide range of DC phenomena that correspond with the latest observations discussed in the review, where roughness effects on DC proved to have a significant role in charge accumulation \cite{aslyamov2022properties}. We get a smoothness of potential near the electrode surface and its enhancement at low positive values of potential near a value of the potential of zero charge (PZC) that was previously obtained in molecular dynamics (MD) numerically \cite{xing2012nanopatterning} and theoretical study \cite{aslyamov2021electrolyte}. Besides, PZC value growth in magnitude with an increase of cation size, temperature, and surface roughness was shown that coincides with electrocapillary measurements \cite{alam2007measurements}, impedance spectrometry \cite{costa2010double}, and MD simulations \cite{vatamanu2011influence}, respectively.

Moreover, we obtain various dependencies of DC over temperature and ion sizes relying on the ions separation model and compared these results with experimental measurements \cite{lockett2008differential, liu2014measurements} and  MD study \cite{vatamanu2011influence}, where we find good qualitative agreement. One of the most interesting effects we describe is the DC profile transition from camel to bell and inverse due to electrode roughness like in MD simulations \cite{vatamanu2011influence}. For the first time, DC profile with more than 2 peaks is obtained analytically. The formation of the third peak in the DC profile occurs due to ion reorientation that we compare with the MD study \cite{xing2012nanopatterning}. We also provide the analysis of parameters, required to enhance DC maxima: IL type and concentration and scale of roughness. We believe that the presented analytical models will help to take a step forward in understanding the molecular mechanisms of the influence of inhomogeneities on the structure and properties of EDL and move forward design of EDLC tools.

The paper has the following structure. The basic equations and definitions are given in section \ref{sec:Theory}, where in the subsections we give solutions for each analytical model: \ref{sec:ISS_model} for the ISS model and \ref{sec:RS_model} for the RS model. The results are presented in section \ref{sec:Results}. In this section, we provide a small review of each considered phenomenon and give a detailed comparison with experimental or numerical studies. At the end of this section, we analyze how to enhance capacitance properties.

\section{Theoretical Model}\label{sec:Theory}

We consider ionic liquid as a binary hard sphere fluid with diameters $d_i$. Let us assume the size ratio $d_2>d_1$ and the charges $Q_1=-Q_2=e$. The solid morphology is defined by the height function $z_s$. In Fig.\ref{fig:TOC}, we schematically illustrate how ions are separated on a rough electrode surface. Accounting steric interactions, small ions can penetrate deeper into the solid surface, so the difference in ion sizes will cause the difference of ion penetration depths $\Delta$, as shown in Fig.\ref{fig:TOC}. Therefore, ion size asymmetry plays a key role in ions separation. Another important parameter that can change $\Delta$ is surface roughness. If we consider a flat surface, then the difference of ion penetration depths is determined by the difference of ion sizes only. However, roughness of electrode surface can enhance ion separations, as happens in Fig.\ref{fig:TOC} on the left. Since ions of IL usually have complex structures, it means that, depending on ions orientation, value of parameter $\Delta$ varies.
In Fig.\ref{fig:TOC} on the right, we show cations that have an elongated shape, which is bigger than anion size from one side. Here, we illustrate how cations are oriented perpendicularly to the surface and penetrate deeper than anions, so the sign of penetration depth parameters is changed if we refer to anion penetration depth. Such case can take place for a negative charge on electrode surface. 

When the Debye length is large, roughness is insensible for ions. As smaller the Debye length as more details of surface structure are captured by ions, so it is the key parameter that makes roughness of electrode surface considerable for DC. Moving to roughness, it is commonly characterized by two parameters: height deviation $\sigma$ and correlation length $l$. We decide to select height deviation as one of the main scale parameters to build a simple 1D analytical model. While correlation length is integrated into the difference of ion penetration depths parameter. This complex parameter includes characters of both roughness parameters, ion sizes, and ion size asymmetry. It captures that surface roughness should be correlated with ion sizes to provide optimal DC characteristics. As bigger difference of ion penetration depths as more ions are separated and this provides higher DC values.

Molecular interactions between solid particles and ions modify ions free energy. The exact free energy term of solid with complex morphology is replaced by effective mean-field free energy term. In our model, we operate with the characteristic function of effective solid potential $S(z)$:
\begin{equation}
    S(z) = \frac{A_{open}(z)}{A_{open}(z) + A_{solid}(z)}
\end{equation}
Then, the effective solid potential expresses as:
\begin{equation}
    U_{eff}(z) = -\frac{1}{\beta} \log S(z)
\end{equation}
$S(z)$ represents Gauss cumulative distribution function for rough surfaces and turns to Heaviside functions for a plane surface. Thus, a shift of solid characteristic functions for particles with different diameters is a difference of particle penetration depths in solid $\Delta$ and the length of a bounce of $S(z)$ characterizes the scale of solid roughness $\sigma$.

As consequence, solid potential has an influence on the ion distribution function.Accounting for solid geometry, the ions density distributions are defined as \cite{aslyamov2021electrolyte}: 
\begin{align}
\label{eq:ions_distr_1}
\rho_1(z)=ec_0 S_1(z)\frac{e^{-\beta e \phi}}{1 + \sum_i \gamma_i \left[e^{-Z_i \beta e \phi} - 1\right]} \\
\rho_2(z)=-ec_0 S_2(z)\frac{e^{\beta e \phi}}{1 + \sum_i \gamma_i \left[ e^{-Z_i \beta e \phi} - 1\right]}
\end{align}
where $\phi$ is the external electrostatic potential, $c_0$ and $\gamma_i$ are bulk ion concentration and ion compacity parameters, respectively. 

Further, the parameters $\Delta$ and $\sigma$ determine the character of ion separation process. There are distinguished two cases, where charge separation occurs due to the difference of ion diameters on a surface with small scale roughness, much smaller than molecular size of ions ($\sigma \ll \Delta$), or the difference of ion diameters and big scale roughness of surface ($\sigma \gg \Delta$).
We denote these cases as ion size separation (ISS) for $\sigma \ll \Delta$ and roughness separation (RS) for $\sigma \gg \Delta$.

\subsection{Ion separation induced by difference of ion sizes}\label{sec:ISS_model}

\begin{figure}[h]
    \centering
    \includegraphics[width=0.75\linewidth]{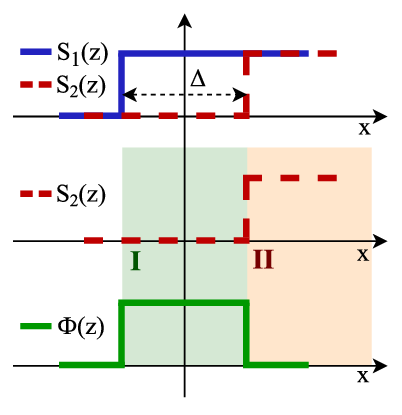}
    \caption{Characteristic functions of effective solid potential for ions on small scale roughness surface.}
    \label{fig:S_first_case}
\end{figure}

In the case of ion size separation $(\sigma \ll \Delta)$, we consider plane solid and different diameters of IL cation and anion. Let us assume, $\Delta = d_2 - d_1$, where $d_1$~--- diameter of positive charges and $d_2$~--- of negative one. Characteristic functions $S_1$ and $S_2$ are taken similar to Heaviside functions and are shown in Fig.\ref{fig:S_first_case}. We assume:
\begin{equation}\label{eq:4}
    S_1(z) = S_2(z) + \Phi(z)
\end{equation}

Expressions for charge distribution in areas $I$ and $II$, illustrated in Fig.\ref{fig:S_first_case}, are given by:
\begin{equation}\label{eq:ions_distr_11}
     \rho^{I}(z)=ec_0 \Phi(z)\frac{e^{-\beta e \phi}}{1 + \sum_i \gamma_i \left[e^{-Z_i \beta e \phi} - 1\right]} \\
\end{equation}

\begin{equation}\label{eq:ions_distr_12}
    \rho^{II}(z)=-2ec_0 S_2(z) \frac{\sinh\left(  \beta e \phi\right)}{1 +2\gamma \sinh^2\left( \beta e \phi /2\right)}
\end{equation}
here, we keep the denominator in r.h.s. of Eq.\ref{eq:ions_distr_12} as in the Kornyshev model \cite{kornyshev2007double} to get through to the analytical solution. Actually, it will not have a serious impact on the solution as it will be simplified for small potentials. 

Further, we derive analytical expressions for cumulative charge and differential capacitance of this system. If we substitute Eqs.\ref{eq:ions_distr_11}-\ref{eq:ions_distr_12} in the Poisson equation, we obtain systems \ref{eq:problem_11}-\ref{eq:problem_12}. To simplify equations, new dimensionless variables are denoted:
$u = \beta e \phi$, $V = \beta e E$, and $x = z/L_D$, where $L_D$ is the Debye length. We account that $\Phi = 1$ in the first area, and $S_2 = 1$ in the second area. Substituting this in Eqs.\ref{eq:ions_distr_11}-\ref{eq:ions_distr_12}, we obtain the following system to solve in the first area $x \in \left[0, \Delta/L_D \right)$:
\begin{equation}\label{eq:problem_11}
\left\{
\begin{array}{rcl}
\frac{\partial^2 u^I}{\partial x^2} &=& -\frac{1}{2}\frac{e^{-u^I}}{1 + \sum_i \gamma_i \left[ e^{-Z_i u^I} - 1\right]}\\
    u^I|_{x=0} &=& V
\end{array}
\right.
\end{equation}
In the second area $x \in \left[\Delta/L_D, +\infty \right)$, it takes the form:
\begin{equation}\label{eq:problem_12}
\left\{
\begin{array}{rcl}
\frac{\partial^2 u^{II}}{\partial x^2} &=& \frac{\sinh(u^{II})}{1 + 2\gamma \sinh^2(u^{II}/2)}\\
    u^{II}|_{\frac{\Delta}{L_D}} &=& u^{I}|_{\frac{\Delta}{L_D}} \\
    u_x^{II}|_{\frac{\Delta}{L_D}} &=& u_x^{I}|_{\frac{\Delta}{L_D}}
\end{array}
\right.
\end{equation}

To solve Eq.\ref{eq:problem_11}, we suggest a small $u$ due to a high temperatures of the system and low particle density. Then, we can follow the classical Debye\,--\,Huckel theory and use linearization to obtain the solution:
\begin{equation}\label{eq:solution_1_1} 
    u^I(x) = \frac{1}{2\lambda^2} + C_1 e^{\lambda x} + C_2 e^{- \lambda x}
\end{equation}
with $\lambda^2 = (1-\gamma_1 + \gamma_2)/2$ and boundary condition: $\frac{1}{2\lambda^2} + C_1 + C_2 = V$.

In the second area, we do not need to solve the system, it will be enough to find the first derivative of $u(x)$, which was obtained in \cite{kornyshev2007double}:
\begin{equation}
    \frac{\partial u^{II}}{\partial x} = \mp \sqrt{\frac{2}{\gamma}} \sqrt{\log\left(1+2\gamma \sinh^2\left(\frac{u^{II}}{2}\right)\right)}
\end{equation} 
To achieve analytical formulation, the Kornyshev model is simplified to Gouy--Chapman approximation for small $u$.

The final system to solve is:
$$\left\{
    \begin{aligned}\label{eq:system_11}
    1/2\lambda^2 + C_1 &+ C_2 = V\\
    \left.\frac{\partial u^I}{\partial x} \right|_{x=\frac{\Delta}{L_D}} &= \left.\frac{\partial u^{II}}{\partial x} \right|_{x=\frac{\Delta}{L_D}}
    \end{aligned} \right.$$
Accounting that $u^{II}|_{\frac{\Delta}{L_D}} = u^{I}|_{\frac{\Delta}{L_D}}$ and suggesting $\Delta \ll L_D$, we obtain the expression for the potential:
\begin{equation}\label{eq:el_pot_model_1}
\begin{aligned}
u(x) &= 1/2\lambda^2 + C_1 e^{ \lambda x} + (V-C_1-1/2\lambda^2) e^{- \lambda x}\\
&= 1/2\lambda^2 + (V-1/2\lambda^2) e^{- \lambda x} + 2C_1\sinh(\lambda x)
\end{aligned}
\end{equation}
where $C_1$ takes form:
\begin{equation}
    C_1 = \frac{\frac{1}{2}(V-\frac{1}{2\lambda^2})-\frac{1}{\lambda}\sinh(V/2)}{1+\alpha \cosh(V/2)}
\end{equation}
The detailed derivation is given in Appendix \ref{sec:ISS_DC}.

\begin{figure} [!b]
    \centering
    \includegraphics[width=0.8\linewidth]{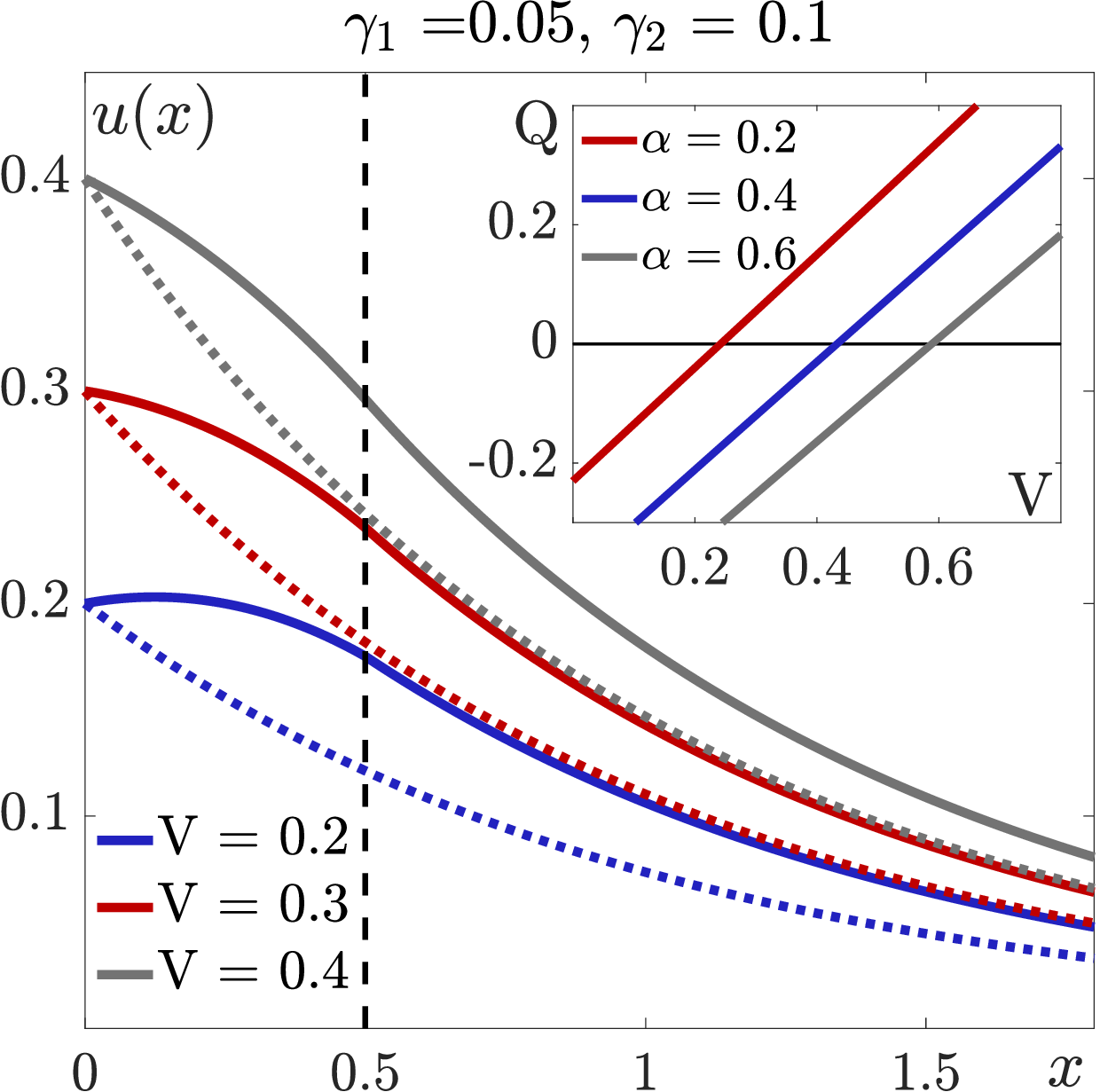}
    \caption{Profiles of electrostatic potential plotted for different potentials on electrode. The dashed line split areas $I$ and $II$ at $\Delta = 0.5 L_D$. In the inset, there is charge dependence on electrode potential for different $\alpha$. }
    \label{fig:potential_first_case}
\end{figure}

The result for electrostatic potential is presented in Fig.\ref{fig:potential_first_case}. It is worth noting that our solution for electrostatic potential describes it in the first area $x \in \left[0, \Delta/L_D \right)$. In the second area, we show the Kornyshev result for electrostatic potential \cite{kornyshev2007double} (see Eq.16). We obtain smooth behavior of $u(x)$ near the electrode surface rather than the asymptotic one, shown in \cite{kornyshev2007double} (see Fig.1). As one can observe, electrostatic potential gradient changes its sign near $x = 0$ with potential growth. Since $Q \sim -\left.\frac{\partial u}{\partial x} \right|_{x=0}$, it means that the potential value at which $\left.\frac{\partial u}{\partial x} \right|_{x=0} = 0$ is the zero charge potential (PZC). At $V<V_{zc}$, the electrostatic potential profile is nonmonotonic that occurs due to the difference of ions penetration depths. At high $V \gg V_{zc}$, it becomes asymptotic like in \cite{kornyshev2007double}. Thus, the obtained result clearly demonstrates the nature of PZC and the transition from the molecular scale effect of ions separation to the common asymptotic behavior of electrostatic potential. This result also represents $u(x)$ local increasing near the electrode surface due to cations presence near the electrode surface for law positive potential values, $u(x)$ profiles without this effect are shown with dashed lines. This effect was previously described in the works \cite{xing2012nanopatterning, aslyamov2021electrolyte}. 

Now, we can find cumulative charge $Q$ formed by the double electric layer structure (it is equal to surface charge due to electroneutrality condition):
\begin{equation}
\label{eq:charge}
Q=\int dz(\rho_1(z)-\rho_2(z))
\end{equation}
It is known that $Q \sim -\left.\frac{\partial u}{\partial x} \right|_{x=0}$ due to the Gauss law, thus:
\begin{equation}\label{eq:Q}
    Q \sim \frac{\alpha\lambda(V-1/2\lambda^2)\cosh(V/2) + 2\sinh(V/2)}{1 + \alpha\cosh(V/2)}
\end{equation}
In the inset of Fig.\ref{fig:potential_first_case}, we show charge-potential dependence. Here, the dimensionless charge should be multiplied by $\frac{\epsilon}{4 \pi e \beta L_D}$. The cumulative charge has negative values at negative potentials and small positive potentials till PZC. It explains the presence of cations near the electrode surface at a small positive value of $V$. We observe that $Q$ shifts to positive PZC values and the shift becomes smaller with $\alpha$ value growth. That brings the idea that there is a limit for $V_{zc}$ value.

Accepting $Q=0$ and small $V$, we obtain the expression for zero charge potential:
\begin{equation}
\label{eq:pzc_1}
    V_{zc} = \frac{\alpha}{2\lambda (1 + \alpha \lambda)}
\end{equation}
In the limit of $\alpha \to \infty$, PZC converges to $1/2\lambda^2$.

Next, we derive the differential capacitance of EDL. The exact expression is given in appendix Eq.\ref{eq:capacity_1}.
\begin{equation}
\label{eq:differential_capacitance}
    C_D=\frac{\partial Q}{\partial E}=
    \left.-\frac{\epsilon}{4\pi L_D}\frac{\partial}{\partial V}\frac{\partial u(x)}{\partial x}\right|_{x=0}
\end{equation}

\subsection{Ion separation induced by electrode roughness} \label{sec:RS_model}

\begin{figure}
\centering
\includegraphics[width=1\linewidth]{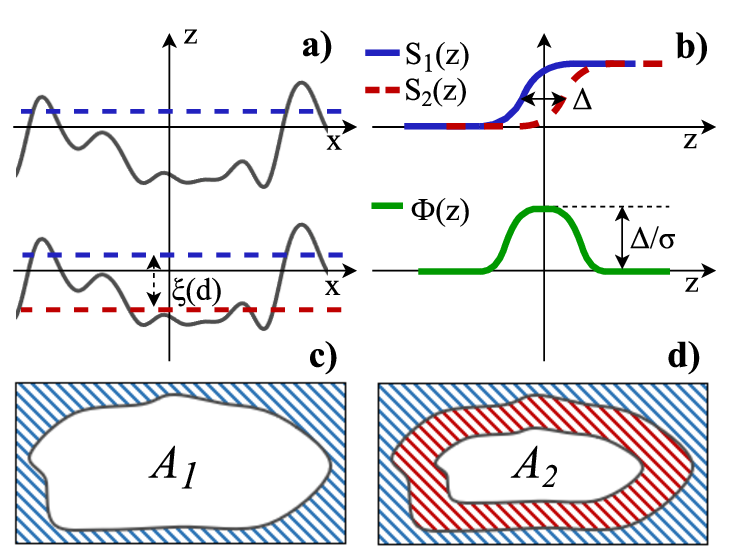}
\caption{(a)~--- Shift of characteristic function of effective solid potential on rough surface, (b)~--- Characteristic functions of effective solid potential for ions on rough surface, (c)~--- Schematic illustration of area available for small cations, (d)~--- Schematic illustration of area available for big anions.}
\label{fig:S_second_case}
\end{figure}

In the case of roughness separation $(\sigma \gg \Delta)$, we study EDL near a significantly rough surface with a small difference in IL ions diameters. If we consider a level and appropriate cut of rough surface, bigger ions will have a smaller available area (see Fig.\ref{fig:S_second_case}(c,d)). To account for particle size, we shift $S(z)$ on a value, that depends on particle diameter $S(z - \xi(d))$ (see Fig.\ref{fig:S_second_case}(a)).

Thus, $S_1(z) = S(z - \xi(d_1))$ and $S_2(z) = S(z - \xi(d_2))$. Accounting that $\Delta = \xi(d_2) - \xi(d_1)$ is small, one can obtain $\Phi(z)$ as:
\begin{equation}
 \Phi(z) = S(z) - S(z-\Delta) =  S'(z) \Delta
\end{equation}
In this case, $S(z)$ is the cumulative Gauss distribution function, and the width of the perturbation area is about $\sigma$, then $\Phi(z)$ equals:
\begin{equation}
 \Phi(z) = \frac{1}{\sqrt{2\pi}} \frac{\Delta}{\sigma} e^{-\frac {z^2}{2\sigma^2}} = \mu g(z)
\end{equation}
where $\mu = \Delta / \sigma \ll 1$ and $g(z) = \frac{1}{\sqrt{2\pi}} e^{-\frac {z^2}{2\sigma^2}}$. Characteristic functions of effective solid potential and perturbation, induced by surface roughness, are shown in Fig.\ref{fig:S_second_case}(b). Further, we consider $S_1(z)$ as the Heaviside function to solve the problem analytically. This transformation will not descend the theory as we are interested only in perturbation influence that is not changed.

To obtain the expression for ion distribution, we use $S_2(z) = S_1(z) - \Phi(z)$. Ion distribution in this case has two terms: the main term and the perturbation term,
\begin{equation}\label{eq:ions_distr_2}
\begin{aligned}
    \rho(z)=&-2ec_0 S_1(z) \frac{\sinh\left( \beta e \phi\right)}{1 + \sum_i \gamma_i \left[e^{-Z_i \beta e \phi} - 1\right]}\\ & + ec_0 \Phi(z)\frac{ e^{ \beta e \phi}}{1 + \sum_i \gamma_i \left[e^{-Z_i \beta e \phi} - 1\right]}
\end{aligned}
\end{equation}

Substituting ion distribution from Eq.\ref{eq:ions_distr_2} into the Poisson equation with dimensionless variables, we obtain the system, where in r.h.s. of the equation on $u$ will be the sum of Kornyshev and perturbation terms. We solve the problem only for $z>0$.
$$\left\{
    \begin{aligned}\label{eq:problem2}
    \frac{\partial^2 u}{\partial x^2} = &\frac{\sinh(u)}{1 + \sum_i \gamma_i \left[ e^{-Z_i u^I} - 1\right]} - \frac{\mu}{2}\frac{\tilde g(x)e^{u}}{1 + \sum_i \gamma_i \left[ e^{-Z_i u^I} - 1\right]}\\
    &\left.u \right|_{x=0} = V
    \end{aligned} \right.$$
Let's consider solution in the following form:
\begin{equation}\label{eq:pr2_sol_type}
    u = u_0 + \sum_{k = 1}^{\infty} \mu^k u_k
\end{equation}
with restrictions:
\begin{equation}\label{eq:problem_2_restrictions}
\begin{array}{lccll}
u_0(x=0) = V& & &\lim_{x \to \infty} u_0 = 0\\
u_k(x=0) = 0& & &\lim_{x \to \infty} u_k = 0, & \forall k
\end{array}
\end{equation}
Substituting $u$ in the system and splitting the problem for each term, we obtain two systems to solve for $u_0$ (Kornyshev solution) and system for $u_1$ (perturbation): 
\begin{equation}\label{eq:problem2_0}
\left\{
\begin{array}{lll}
\Delta u_0 =& \frac{\sinh(u_0)}{1 + \sum_i \gamma_i \left[ e^{-Z_i u_0} - 1\right]}\\
    \left.u_0 \right|_{x=0} &=V&  \\
    \left.u_0 \right|_{x \to \infty} &=0& 
\end{array}
\right.
\end{equation}

\begin{equation}\label{eq:problem2_1}
\left\{
\begin{array}{lll}
\Delta u_1 =& \left.\frac{\partial f}{\partial u}\right|_{u_0} u_1 &- \frac{1}{2} \tilde g(x) e^{u_0}\\
    \left.u_1 \right|_{x=0} &=0&  \\
    \left.u_1 \right|_{x \to \infty} &=0& 
\end{array}
\right.
\end{equation}
For small $\gamma_i$, r.h.s of first equation in system \ref{eq:problem2_0} turns to $f(u_0) = \sinh(u_0)(1 + \gamma_1 + \gamma_2 - \gamma_1 e^{-u_0} - \gamma_2 e^{u_0})$. We denote:
\begin{equation}\label{eq:lambda}
    \lambda^2 = \left.\frac{\partial f}{\partial u}\right|_{u_0} = \gamma_1 + \gamma_2 + \cosh(u_0)(1 + \gamma_1 + \gamma_2)
\end{equation}

The first equation in system \ref{eq:problem2_1} turns into screened Poisson equation that can be solved using the Green function. But because of boundary conditions, we are to use the "method of images" to find the Green function.

Then, the solution $u_1$ is expressed as:
\begin{equation}
    u_1(x) = \frac{1}{4\lambda} \int_0^{\infty} \tilde g(x_0) e^{u_0(x_0)}\left[e^{-\lambda |x-x_0|} - e^{-\lambda |x+x_0|}\right] d x_0
\end{equation}

\begin{figure}
    \centering
    \includegraphics[width=0.8\linewidth]{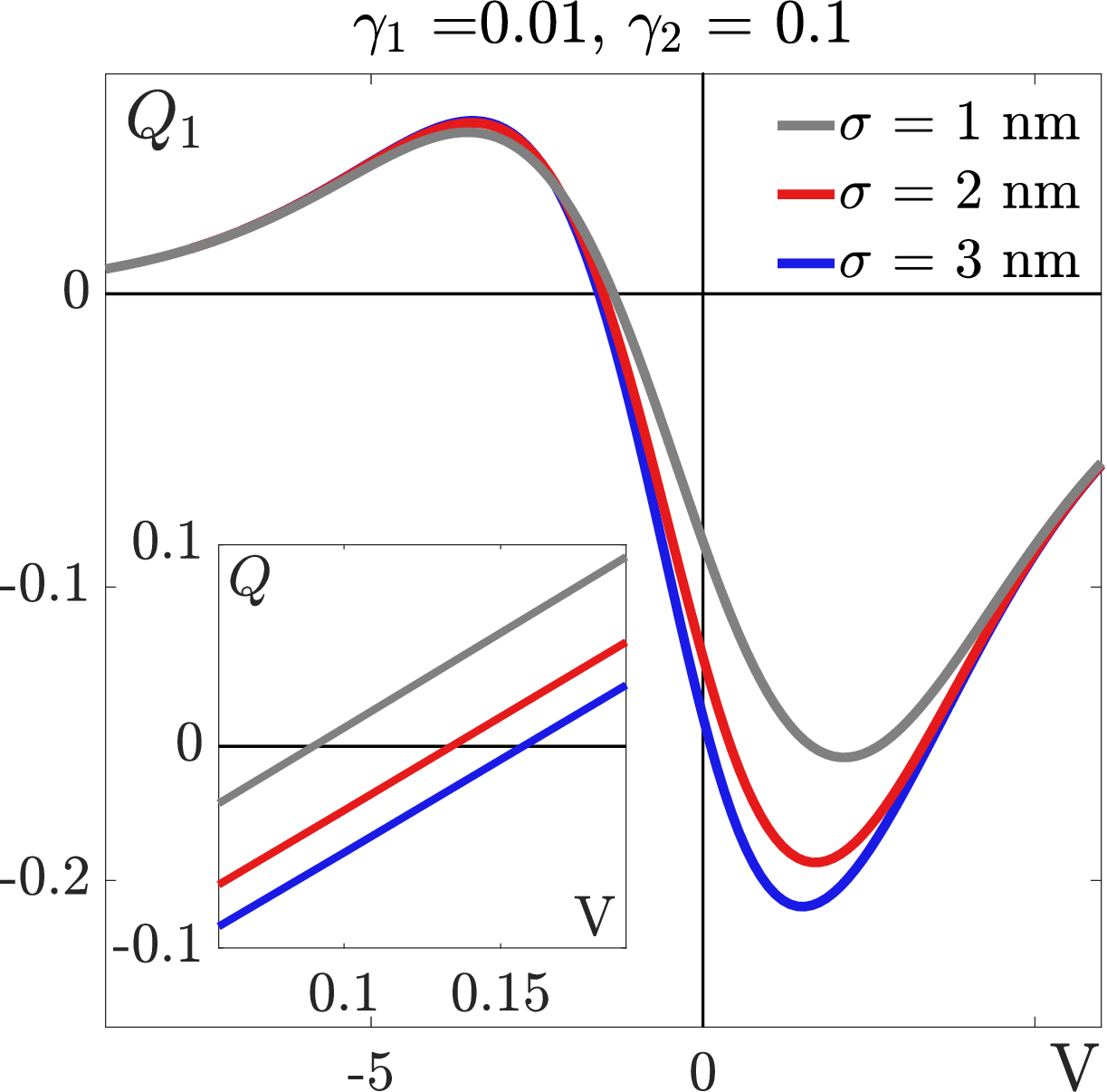}
    \caption{Charge perturbation plotted for different electrode roughness parameters. In the inset, there is charge dependence on electrode potential for these roughness parameters.}
    \label{fig:Q_1_second_case}
\end{figure}

Next, we find cumulative charge on electrode $Q$, accounting Eq.\ref{eq:pr2_sol_type}.
\begin{equation} \label{eq:charge_2}
Q \sim -\left.\frac{\partial u}{\partial x} \right|_{x=0} = -\left.\frac{\partial u_0}{\partial x}\right|_{x=0} - \mu \left.\frac{\partial u_1}{\partial x}\right|_{x=0}
\end{equation}
here, the first term is Kornyshev charge $Q_0 \sim 2\sinh(V/2)$. Charge perturbation is given by:
\begin{equation} \label{eq:charge_petrub_2_com}
Q_1 \sim - \mu \left.\frac{\partial u_1}{\partial x}\right|_{x=0} =
 -\frac{\Delta}{2 \sigma} \int_0^{\infty} \tilde g(x_0) e^{u_0(x_0)}e^{-\lambda x_0} d x_0
\end{equation}
Using the Kornyshev solution for $u_0$, implying a small potential assumption and Laplace transform of the Gaussian curve, we obtain the final expression for charge perturbation:
\begin{equation} \label{eq:charge_perturb}
\begin{aligned}
    Q_1 \sim &-\frac{\Delta}{4L_D} \left[e^{\frac{\lambda^2 \sigma^2}{2 L_D^2}} \erfc\left(\frac{\sqrt{2}\lambda}{2} \frac{\sigma}{L_D}\right)  + \right. \\&\left.4\tanh\left(\frac{V}{4}\right)e^{\frac{(\lambda + 1)^2 \sigma^2}{2 L_D^2}} \erfc\left(\frac{\sqrt{2}(\lambda+1)}{2} \frac{\sigma}{L_D}\right) \right]
\end{aligned}
\end{equation}
All details are given in Appendix \ref{sec:RS_DC}, where we also provide approximations for $Q_1$ in cases of large and small Debye lengths. The result for charge perturbation is presented in Fig.\ref{fig:Q_1_second_case}. Its values grow absolutely with $\sigma$ increase. Here, we keep $\mu = 1$ and take $L_D$ = 2 nm. Charge perturbation has positive values for negative potentials and changes sign near $V = 0$. In the inset, we show cumulative charge, where $Q = Q_0 + Q_1$. Observed PZC values are positive and nonmonotonically increase with $\sigma$ growth, so we suggest that there is a limit for $V_{zc}$.

Zero charge potential for the RS model takes the form:
\begin{equation}\label{eq:pzc_2}
      V_{zc} =  \frac{\Delta}{4L_D}\frac{e^{\frac{\lambda^2 \sigma^2}{2 L_D^2}} \erfc\left(\frac{\sqrt{2}\lambda}{2} \frac{\sigma}{L_D}\right)}{1 - \frac{\Delta}{4L_D}e^{\frac{(\lambda+1)^2 \sigma^2}{2 L_D^2}} \erfc\left(\frac{\sqrt{2}(\lambda+1)}{2} \frac{\sigma}{L_D}\right)}
\end{equation}
where $\lambda^2 = 1+ 2\gamma_1 + 2\gamma_2$. In the limit of $L_D \to 0$, it turns to $V_{zc} = [\lambda (2\sqrt{2 \pi} \sigma/\Delta - \frac{1}{\lambda+1})]^{-1}$.

Now, we can find the derivation of $Q_1$ on $E$ and get the expression for differential capacitance perturbation $C_D^1$.

\begin{equation} \label{eq:diff_cap_petrub_2_com}
C_D^1 = \frac{\partial Q}{\partial E} \sim
 -\frac{\Delta}{2 \sigma} \int_0^{\infty} \tilde g(x_0) e^{u_0(x_0)}e^{-\lambda x_0} \left[ \frac{\partial u_0}{\partial E} - \frac{\partial \lambda}{\partial E}\right] d x_0
\end{equation}

The exact  formula for DC is given in Eq.\ref{eq:capacity_2}. We also obtain approximations for large and small Debye lengths that are given in Eqs.\ref{eq:capacity_2_small_L_D}-\ref{eq:capacity_2_large_L_D}.
\section{Results and Discussion}\label{sec:Results}

\subsection{Zero charge potential}

Experimental data for PZC of EDL formed in ionic liquids is discrepant \cite{alam2007measurements, alam2008capacitance, liu2014measurements}. In these works, the PZC of RTILs with imidazolium based cations from C$_2$ to C$_{10}$ and  tetrafluoroborate BF$_4^-$ anion was investigated. Alam et al. \cite{alam2007measurements} obtained negative values of PZC on Hg electrode that grow in absolute value if the alkyl chain length becomes longer.  In the work \cite{liu2014measurements}, the authors provide data for Ag and Au electrodes. The presented PZC results are negative and become smaller in magnitude with cation size increase for both considered electrodes. The presented results \cite{alam2008capacitance} have as positive as negative values. For example, for electrochemically untreated GC electrode PZC for all these IL are positive and increasing with cation size growth. For Au and Pt electrodes, PZC takes negative values that decrease in magnitude for bigger cations. In this review, we found three various behaviors of PZC with cation size raise. This disagreement likely consists in electrode type.

In our results, PZC is described by Eq.\ref{eq:pzc_1} for the ISS  model and by Eq.\ref{eq:pzc_2} for the RS model. In both cases, we consider IL with bigger anions and obtained positive PZC values. To compare our results with data from the literature, we need to change the initial assumption of smaller cations. We provide this correction for the ISS model. Solving the nonlinear Poisson-Boltzmann equation in the new assumption of bigger cations, we derive the following expression for PZC: $V_{zc} = -\frac{\alpha}{2\lambda (1 + \alpha \lambda)}$ with the limit at $\alpha \to \infty$ equal $-1/2\lambda^2$ and $\lambda^2 = (1 + \gamma_1 - \gamma_2)/2$. When we assume bigger cations, our analytical models provide negative PZC values that grow in absolute value with cation size increase.
It coincides with the experimental data from \cite{alam2007measurements}. The sign of PZC for IL with bigger cations also agrees with the results from \cite{liu2014measurements}, for Pt and Au electrodes from \cite{alam2008capacitance}, and for  1-butyl-3-methylimidazolium BMIM$^+$ and  1-hexyl-3-methylimidazolium HMIM$^+$ cations results from \cite{costa2010double}. While we are unable to represent the whole variety of PZC behavior. It seems to suggest that a more complex concept is required to describe PZC specificity.

\begin{figure}
    \centering
    \includegraphics[width=0.9\linewidth]{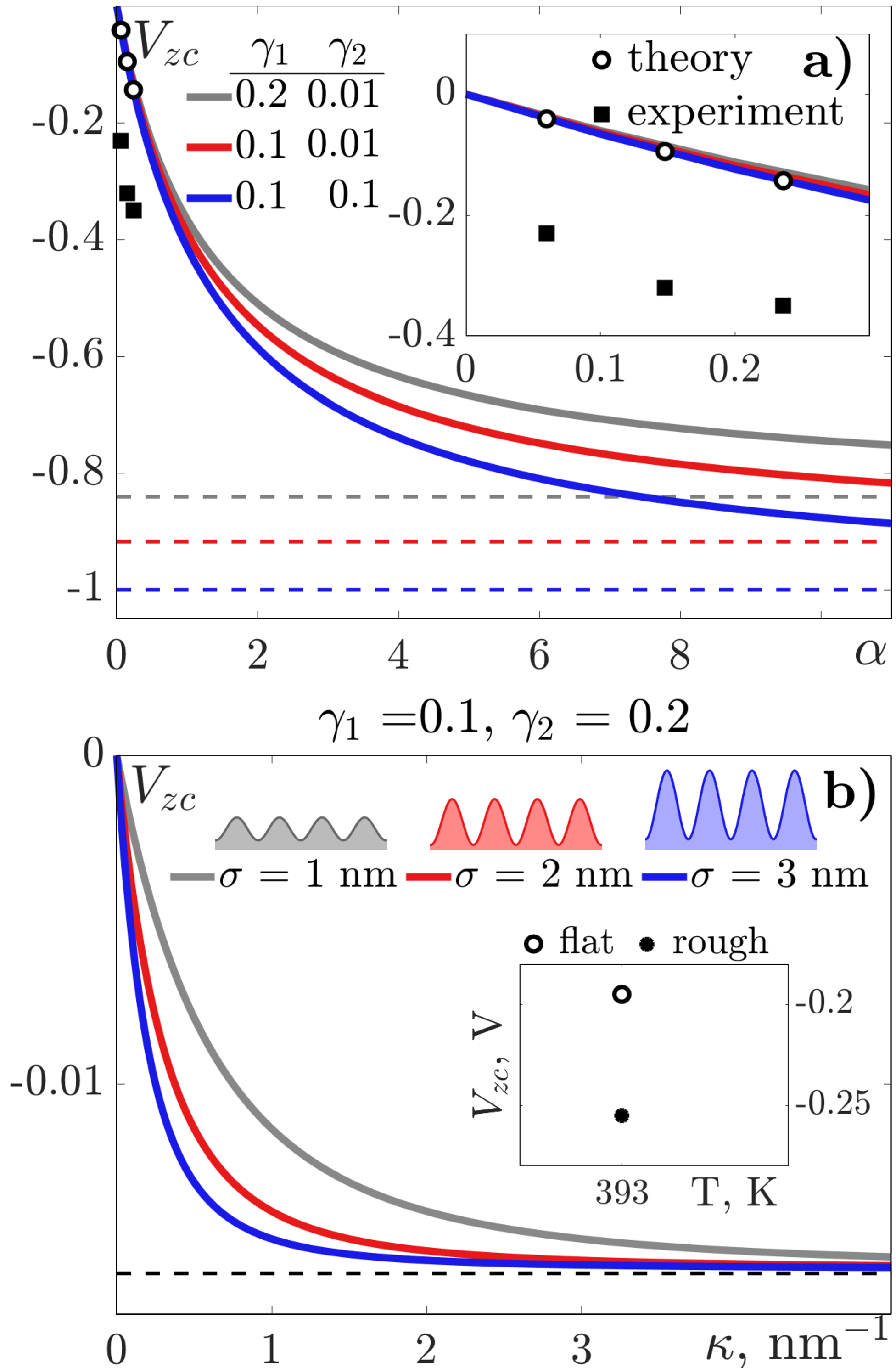}
    \caption{a) Zero charge potential from the IS model for different $\gamma_1$ and $\gamma_2$. In the inset, we plot PZC experimental data from \cite{alam2007measurements} (squares) in comparison with the ISS model results (circles) for considered IL calculated with assumption of $L_D = 5$ nm; b) Zero charge potential from the RS model for various roughness parameters. Inset shows result from MD simulation \cite{vatamanu2011influence} for flat and rough surfaces of electrode.}
    \label{fig:V_zc}
\end{figure}

\subsubsection{Comparison with experimental measurements by Alam et al. \cite{alam2007measurements} and Costa et al. \cite{costa2010double}}

In Fig.\ref{fig:V_zc}(a), we show the result of the ISS model for IL with bigger cations and put electrocapillary measurements from \cite{alam2007measurements}. For calculations, we consider T = 298 K, $L_D$ = 5 nm, and ion diameters 0.46 nm for BF$_4^-$ anion and 0.76 nm, 1.2 nm, 1.64 nm for EMIM$^+$, BMIM$^+$, and HMIM$^+$ respectively. Important to note that our dimensionless $V_{zc}$ should be multiplied on $1/e\beta \approx 1/40 V$ to become a proper value of potential. We obtain that the PZC value grows in magnitude if the cations size increase. The same tendency was observed by Alam et al \cite{alam2007measurements}. Thus, the model provides qualitative relation of PZC on ion sizes, but values don't coincide. 

In addition, PZC dependence on $\alpha$ (see Fig.\ref{fig:V_zc}(a)) can be interpreted as a dependence on temperature. Since $\alpha = \Delta/L_D$, if we keep $\Delta$ constant, it becomes dependent on $\kappa \sim T^{-1/2}$. Therefore, we obtain that with temperature growth PZC value decreases. It coincides with the results of electrochemical impedance spectrometry \cite{costa2010double}. Here, the authors study EMIM$^+$, BMIM$^+$, and HMIM$^+$ with bis(trifluoromethanesulfonyl)imide TFSI$^-$ ionic liquids. TFSI$^-$ ion size is about 1.132 nm according to \cite{sillars2011effect}, which is bigger than EMIM$^+$ cation and smaller than BMIM$^+$ and HMIM$^+$ cations. So, these results also reproduce PZC temperature dependence for BMIM$^+$ and HMIM$^+$ cations with TFSI$^-$ on a Hg electrode from \cite{costa2010double}.

\subsubsection{Comparison with molecular simulation by Vatamanu et al. \cite{vatamanu2011influence}}

The roughness effect on PZC is shown in Fig.\ref{fig:V_zc}(b). Here, we present the results of the RS model in the assumption of bigger cations. PZC falls faster with $\kappa$ increase if the roughness parameter is bigger. We keep $\mu = 0.1$ that explains one limit for the illustrated curves with different $\sigma$. PZC approaches a constant value for small Debye length and roughness helps to reach this limit faster. Our results agree with the results of molecular simulation \cite{vatamanu2011influence}, where authors model [1-ethyl-3-methylimidazolium] [bis(fluorosulfonyl)imide] EMIM$^+$FSI$^-$ ionic liquid near flat and rough graphite surfaces. They provide PZC values for both surfaces at T = 393 K (see inset in Fig.\ref{fig:V_zc}(b)). It was found that on a rough surface PZC was lower than on a flat surface. We also observe decreasing in PZC with roughness in our model. Thus, the roughness of an electrode surface allows us to alter the PZC value. 

PZC values were found to vary with the size of IL ions, temperature, and electrode surface roughness. Also, we observe that PZC reaches limit values at high inverse Debye lengths, i.e. high ions density. We obtain that the sigh of PZC value is determined by ion size ratio: negative for bigger cations and positive for bigger anions. We managed to predict the observed tendencies, but not all of them. Thus, PZC behavior should be further investigated and discussed.
 
\subsection{Differential capacitance dependence on temperature}

DC dependence on temperature was previously studied in the works \cite{ lockett2008differential, silva2008electrical, vatamanu2010molecular, liu2014temperature, chen2018temperature}. However, there is no coherent view of this relation, as both positive and negative DC gradient with temperature was observed. In the works \cite{lockett2008differential, silva2008electrical, liu2014temperature}, it was found that DC raises with temperature. Such behavior is enforced by a weakening of ion association, consequently, more ions become able to get closer to an electrode surface. In the other works \cite{vatamanu2010molecular, chen2018temperature}, DC reduced with temperature growth. This result is usually explained by an increasing role of entropic effects, leading to structural disorder and desorption of charged ion parts. Below, we provide our explanation for the presented inconsistency of temperature effect on DC. 

\subsubsection{Comparison with experimental measurements by Lockett et al. \cite{lockett2008differential}}

\textit{Ion size separation ($\sigma \ll \Delta$).} To study DC dependence on temperature in the case of ion size separation, we consider HMIM$^+$Cl$^-$ ionic liquid and calculate DC by Eq.\ref{eq:capacity_1} for three temperatures 353 K, 373 K, and 393K. The diameter of HMIM$^+$ cation is 1.64 nm and Cl$^-$ ion size is about 0.35 nm according to \cite{levitzki1974allosteric}. We put $\gamma_2 = 10^{-6}$ and calculate $\gamma_1 = \gamma_2 (d_1/d_2)^3$. Obtained dimensionless DC values grow with temperature increase. This result agrees with impedance spectrometry measurements of DC for HMIM$^+$Cl$^-$ on glassy carbon \cite{lockett2008differential}. In this study, three ionic liquids with different cations were examined, and the decreasing tendency was observed for all these ILs.

\begin{figure}
    \centering
    \includegraphics[width=0.9
    \linewidth]{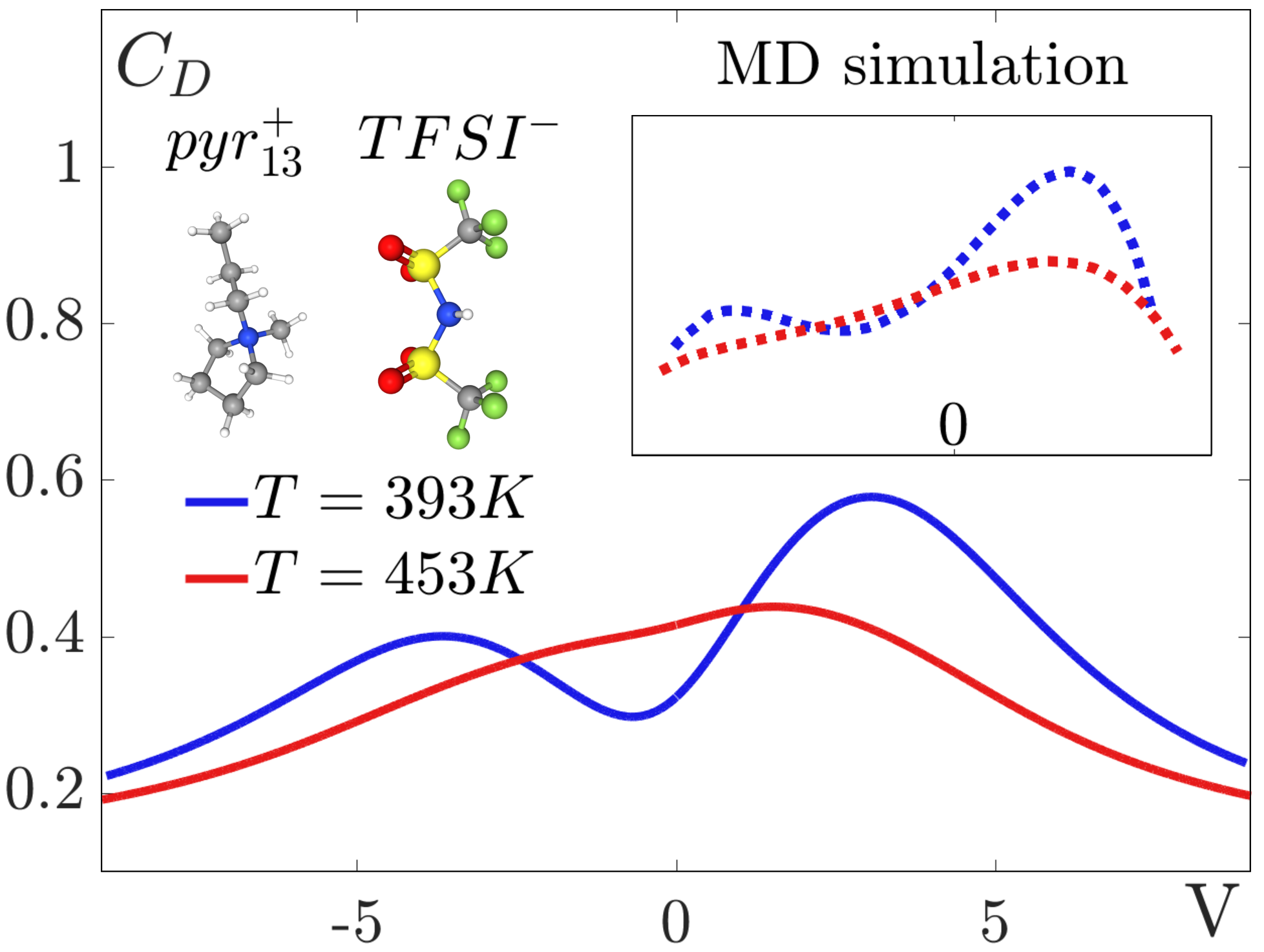}
    \caption{DC profiles for for pyr$_{13}^+$ TFSI$^-$ at different temperatures calculated by RS model. MD simulation data for pyr$_{13}^+$TFSI$^-$ from \cite{vatamanu2010molecular} is plotted in the inset.}
    \label{fig:DC_with_T}
\end{figure}

\subsubsection{Comparison with molecular simulation by Vatamanu et al. \cite{vatamanu2010molecular}}

\textit{Roughness separation ($\sigma \gg \Delta$).}
In Fig.\ref{fig:DC_with_T}(b), we plot DC profiles calculated by Eq.\ref{eq:capacity_2} for [N-methyl-N-propylpyrrolidinium] [bis(trifluoromethanesulfonyl)imide] pyr$_{13}^+$ TFSI$^-$ and two temperatures T = 393 K and T = 453 K. We estimate pyr$_{13}^+$ cation size as EMIM$^+$ cation size $d_1 =$ 0.76 nm, and take TFSI$^-$ ion size, as we mentioned earlier, $d_2 =$ 1.132 nm. We take $L_D$ about 2 nm and account for its temperature dependence. Besides, $\Delta$ changes with temperature, so we consider $\mu = 2$ for T = 453 K and $\mu = 6$ for T = 393 K. Roughness parameter is taken $\sigma$ = 1 nm. Ions compacities were accounted in the following way: $\gamma_1  = 0.1, \gamma_2 = \gamma_1(d_2/d_1)^3$. Presented DC values should be multiplied by factor $\frac{\epsilon}{4 \pi 10^{-9}}$ to become dimensional. DC values change nonmonotonically if temperature grows. The minimum value for lower temperature intersects the DC profile for the higher one. We compare it with the results of MD simulation \cite{vatamanu2010molecular}, where DC of pyr$_{13}^+$TFSI$^-$ on graphite electrode was investigated. Vatamanu et al. also obtained interesting behavior of DC profiles with temperature raise. For higher temperature, they observe smaller DC values and smaller DC variation with potential. They explain the disappearance of DC minima at high temperature by an entropic penalty that IL pays to structure near an electrode surface. Similar results were observed in the works \cite{ vatamanu2014influence, chen2018temperature}. In the work \cite{chen2018temperature}, DC temperature dependence was studied with both
theoretical and computational methods. The predicted DC curves transform from a camel to a bell shape with increasing temperature and DC values drop. In MD simulation \cite{vatamanu2014influence}, electrode roughness influence on DC-temperature dependence was studied. As a result, DC reduces with temperature increase on atomically rough surfaces at particular voltages and roughness causes a larger variation of DC.

Our results show that temperature increase can both enhance and decline DC values depending on the separation process and potential value. 

\subsection{Differential capacitance dependence on ion sizes}

\begin{figure} [!b]
    \centering
    \includegraphics[width=0.9\linewidth]{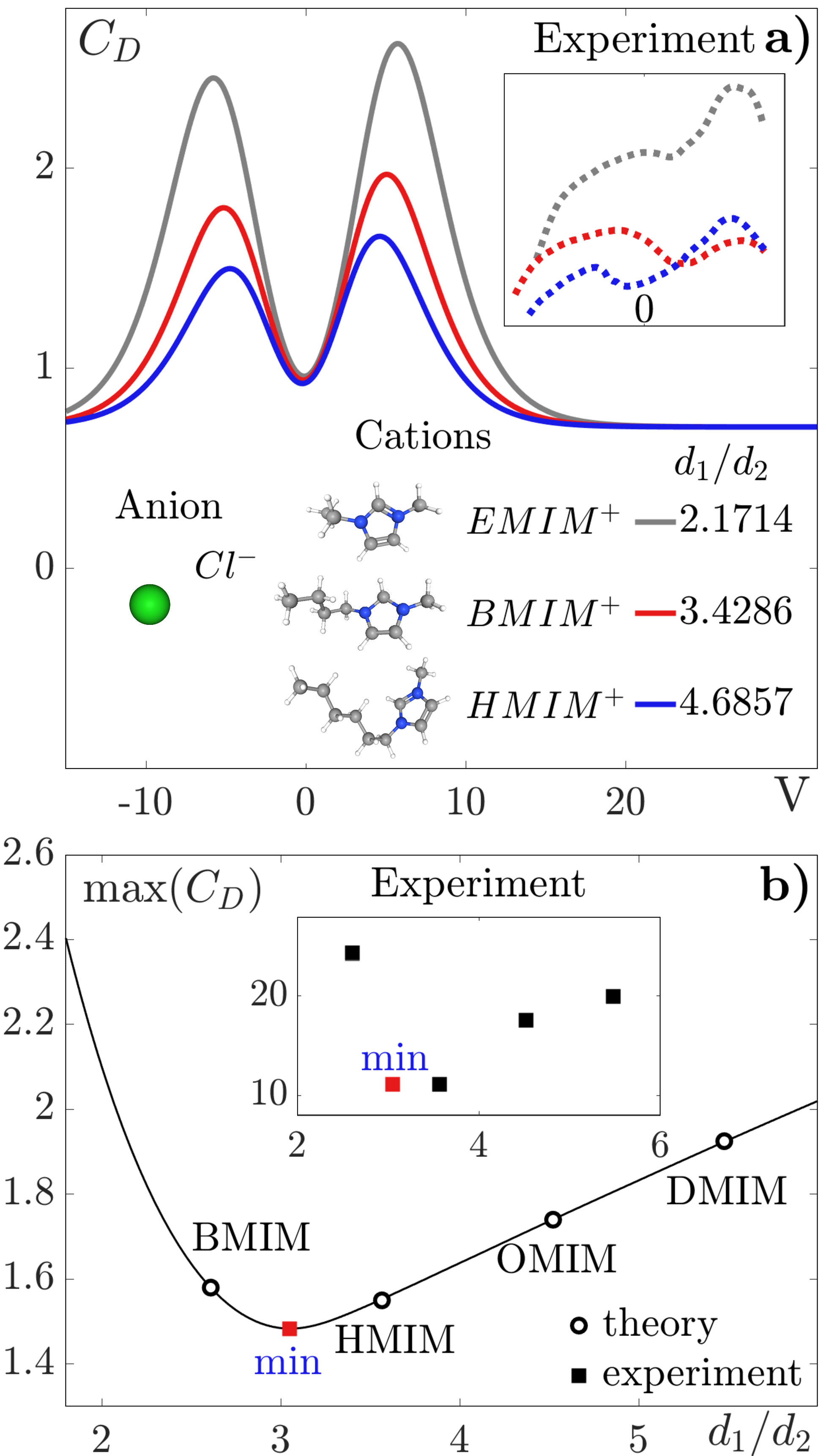}
    \caption{a) DC profile variation with cation size increase for ISS model. In the inset, there is DC experimental data for EMIM$^+$, BMIM$^+$, HMIM$^+$ cations with chloride anion from \cite{lockett2008differential}; b) DC maximum depending on ion sizes relation. Inset shows experimental data for BMIM$^+$, HMIM$^+$, OMIM$^+$, DMIM$^+$ and BF$_4^-$ from \cite{liu2014measurements}.}
    \label{fig:DC_ion_sizes}
\end{figure}

Sizes of IL ions, their relation, and how they are small in comparison with surface roughness play a key role in EDL structure and therefore DC values and form \cite{lockett2008differential,ma2014classical,jo2017effects, liu2014measurements,hu2013molecular}. For example, in MD simulations \cite{hu2013molecular} two ionic liquids BMIM$^+$ PF$_6^-$ (hexafluorophosphate) and BMIM$^+$ BF$_4^-$ with the same cation and two different anions were considered near basal and prismatic graphite and (001) and (011) Au electrodes. As a result, higher max-min changes of DC were observed for more asymmetric IL and a rough electrode surface. Besides, in the works \cite{lockett2008differential,ma2014classical,jo2017effects} EMIM$^+$, BMIM$^+$, HMIM$^+$ ionic liquids were considered with TFSI$^-$ \cite{ma2014classical}, Cl$^-$ \cite{lockett2008differential}, and BF$_4^-$ \cite{jo2017effects} anions near planar \cite{ma2014classical}, glassy carbon \cite{lockett2008differential} and graphene \cite{jo2017effects} electrodes, so in each investigation, anion size was smaller than cations sizes and there was no surface roughness. In these works, DC raises for smaller cations because: (i) EDL becomes thinner, (ii) as longer alkyl chain as more neutral beads it has, and it causes a reduction in electrostatic correlations and decreases effective charge density, (iii) due to steric effects when cation is longer, fewer cations are adsorbed, so DC drops. Another behavior occurred in Raman scattering measurement \cite{liu2014measurements}, where DC maximum has nonmonotonic behavior. They investigated EMIM$^+$, BMIM$^+$, HMIM$^+$, OMIM$^+$ (1-octyl-3-methylimidazolium), and DMIM$^+$ BF$_4^-$ ionic liquids near Ag electrode that may have roughness in several molecular layers. Below, we provide a possible explanation of these effects. 

\subsubsection{Comparison with experimental measurements by Lockett et al. \cite{lockett2008differential}}

 \textit{Ion size separation ($\sigma \ll \Delta$).}
 Impedance spectrometry measurements \cite{lockett2008differential} show that DC values of EMIM$^+$, BMIM$^+$, and HMIM$^+$ Cl$^-$ ionic liquids increase for smaller cations at glassy carbon, i.e. flat electrode. It can be associated with larger specific ion adsorption for smaller cations. We consider these ILs with chloride anion size 0.35 nm and EMIM$^+$, BMIM$^+$, and HMIM$^+$ sizes 0.76 nm, 1.2 nm, and 1.64 nm, respectively. Hence, anion size is smaller than the sizes of cations. We take T = 373 K and $L_D$ = 5 nm. We provide dimensionless values of DC that need to be multiplied by factor $\frac{\epsilon}{4 \pi L_D}$. Results of our analytical solution for the ISS model are presented in Fig.\ref{fig:DC_ion_sizes}(a) and qualitatively coincide with results from \cite{lockett2008differential} that are shown in the inset. Such behavior for bell shaped DC profiles were also obtained in the works \cite{ma2014classical,jo2017effects} with the same cations, but another anion near planar electrodes.
 
 \begin{figure*}
    \centering
    \includegraphics[width=0.9\linewidth]{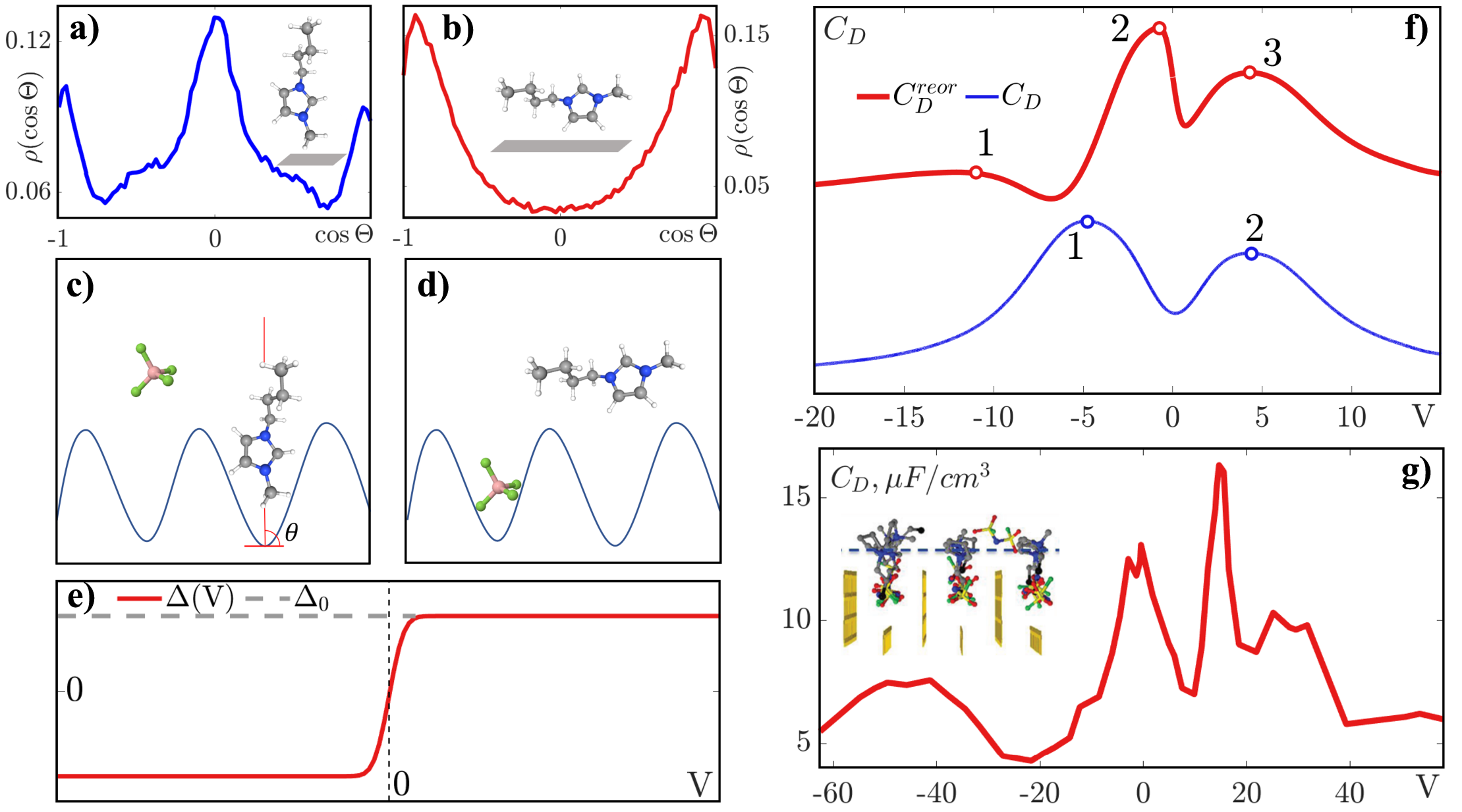}
    \caption{a) Schematic probability distribution of cation orientation at negative potentials;
    b) Schematic probability distribution of cation orientation at positive potentials; c) Schematic representation of cations orientation at negative potential values; d) Schematic representation of cations orientation at positive potential values; e) Penetration depth profile modification due to ion reorientation; f) Formation of third peak in DC profile due to perturbation with ion reorientation (DC with ion reorientation - red line, DC without ion reorientation - blue line); g) DC profile from \cite{xing2012nanopatterning} for S3 surface.}
    \label{fig:reorientation}
\end{figure*}

\subsubsection{Comparison with experimental measurements by Liu et al. \cite{liu2014measurements}}
 
\textit{Roughness separation ($\sigma \gg \Delta$).}
Next, we investigate how DC maximum changes with ion size ratio increase in the RS model. As a result, we find that DC perturbation increase with a growth of ion sizes difference. It is actually obvious, because $\Delta/\sigma$ is the amplitude of perturbation, so as more $\Delta$ or less $\sigma$ as larger perturbation. It can explain the nonmonotonic DC results of surface enhanced Raman scattering \cite{liu2014measurements} by increasing the significance of the perturbation effect with the alkyl length increase. Liu et al. obtained that DC reduces for EMIM$^+$, BMIM$^+$, and HMIM$^+$ cations and grows for HMIM$^+$, OMIM$^+$, and DMIM$^+$ cations. Fig.\ref{fig:DC_ion_sizes}(b) shows how our model represents this nonmonotonic behavior and successfully determines the ion size relation of gradient change. We take $L_D$ = 1 nm, $\sigma$ = 0.1 nm and $\gamma_2 = 0.005$. We work with dimensionless values of DC that should be additionally multiplied by factor $\frac{\epsilon}{4 \pi L_D}$. We obtain that for IL with $d_1/d_2 <3$ perturbation has a lower influence on DC and DC maximum became smaller as it does without perturbation. But, when $d_1/d_2 >3$, the perturbation effect prevails and DC maxima start to grow with cation size increase. In the inset, we give Raman scattering data on DC maxima for IL with BMIM$^+$, HMIM$^+$, OMIM$^+$, and DMIM$^+$ cations from \cite{liu2014measurements}. We can see that the gradient change of the DC maximum obtained by our model catches the behavior of DC data.

DC is found to be strongly dependent on the ion size ratio. In the ISS model, DC grows for a smaller ion size difference. In the RS model, DC undergoes both a decrease and an increase with size ratio growth. The behavior changes till the size ratio reaches a particular value.

\subsection{Ions structural transition effect on differential capacitance}

Orientation of ions in EDL fundamentally determines its structure and therefore behavior of DC. Ion orientation depends on electrode potential, chemical composition of IL ions, and electrode surface structure \cite{vatamanu2010molecular, xing2012nanopatterning, jo2017effects}. Important to note that the orientation of ions in the interfacial layer is of the most interest. 

It was earlier revealed that ions change their orientation in response to electrode potential, and this effect causes alteration in IL dielectric constant near the electrode surface that provides DC peak formation \cite{ vatamanu2010molecular}. In the work \cite{vatamanu2010molecular}, both pyr$_{13}^+$ cations  and TFSI$^-$ anions orient mostly perpendicular to the graphene surface near PZC. In addition, if potential grows, more counterions in the closest layer orient parallel to the surface. While, in the work \cite{jo2017effects}, cations EMIM$^+$, BMIM$^+$, and HMIM$^+$ orient predominantly in parallel to the graphite electrode surface at all considered potentials, and this effect becomes stronger for longer cations. In MD simulation \cite{xing2012nanopatterning}, authors study DC of pyr$_{13}^+$ FSI$^-$ near graphite electrode with various surface roughness. They demonstrated that the appropriate design of nanoroughness can significantly improve the capacitance properties of EDL. For smaller roughness, authors obtained formations of new peaks in the DC profile that was caused by ion reorientation. So, these observations confirm that the orientation of IL ions varies with their chemical composition, potential, and surface structure and can modify DC profile and values.

We investigate the effect of ions reorientation by modifying the perturbation term. To represent reorientation, we turn the penetration depth parameter into a function of potential in the RS model. Previously, we used a constant value of penetration depth $\Delta = \Delta_0$ that was determined by the difference of ion diameters. Now, the penetration depth parameter has potential dependence: $\Delta (V) = \Delta_0 - (\Delta_0+\sigma)\erfc(2V)$. This idea is illustrated in Fig.\ref{fig:reorientation}(e). We consider IL with small anions that penetrate deeper than big cations when potential is positive, which corresponds with $\Delta > 0$. At these conditions, the considered system is in a homogeneous state, and cations orient parallel with the electrode surface (see Figs.\ref{fig:reorientation}(b,d,e)). When an electrode potential becomes negative, it attracts cations, and they change their orientation to a perpendicular one and start to penetrate more than anions. This state is inhomogeneous with $\Delta < 0$ (see Figs.\ref{fig:reorientation}(a,c,e)). Thus, the sign of $\Delta$ changes when IL turns from a homogeneous to an inhomogeneous state. Since this parameter is involved as the amplitude coefficient for DC perturbation, it leads to DC modification. 

\subsubsection{Comparison with molecular simulation by Xing et al. \cite{xing2012nanopatterning}}

The result of the DC profile with ion reorientation is presented in Fig.\ref{fig:reorientation}(f). We take $L_D$ = 10 nm, $\sigma$ = 1 nm, $\Delta_0$ = 15 nm, $\gamma_1$ = 0.01, $\gamma_2$ = 0.2. We provide dimensionless values of DC that need to be multiplied by factor $\frac{\epsilon}{4 \pi L_D}$. We notice that the DC curve without account of ions orientation gives two peaks and, if ion orientation is considered, we obtain three peaks of the DC profile. To our knowledge, it's the first time when an analytical solution provides the formation of more than two peaks in the DC profile. In Fig.\ref{fig:reorientation}(g), we plot the DC result obtained by MD simulation \cite{xing2012nanopatterning}. Here, pyr$_{13}^+$FSI$^-$ ionic liquid was simulated in contact with 3 types of the patterned electrode surface. S3 is the designation of the surface with the smallest scale of corrugation: correlation length is 0.73 nm and height deviation is 0.513 nm. Xing et al. explained the formation of new peaks in the DC profile by ion reorientation and showed that DC is strongly dependent on the surface structure. The additional peak for negative potentials obtained in our model is similar to the narrow peak at $V \in [-2, -1]$ for DC on the S3 electrode.

We provide RS model modification that treats the ion reorientation effect. This model managed to reproduce the formation of the additional peak in the DC curve, so it expands bell and camel-like types of DC analytical solutions.

\subsection{Transition of DC-curves from
camel to bell shapes}

The transition of the DC profile from camel shape to bell shape due to ions concentration is well known \cite{kornyshev2007double, jiang2011density, fedorov2014ionic}. However, in MD simulations \cite{vatamanu2011influence, hu2014comparative}, other reasons for this effect were found like surface roughness or anion size increase. Roughness induced transition can be confirmed by electrochemical impedance measurements \cite{islam2008electrical}, where researchers obtained bell shaped DC for Au and Pt electrodes and U-shaped for glassy carbon. This effect was also discussed in the review \cite{aslyamov2022properties}. Meanwhile, in MD studies \cite{vatamanu2012molecular, hu2013molecular}, where IL was considered near flat and rough surfaces, DC form does not change in this way. So, we can decide that more complex conditions are required to create this transition.

\begin{figure}
    \centering
    \includegraphics[width=0.9\linewidth]{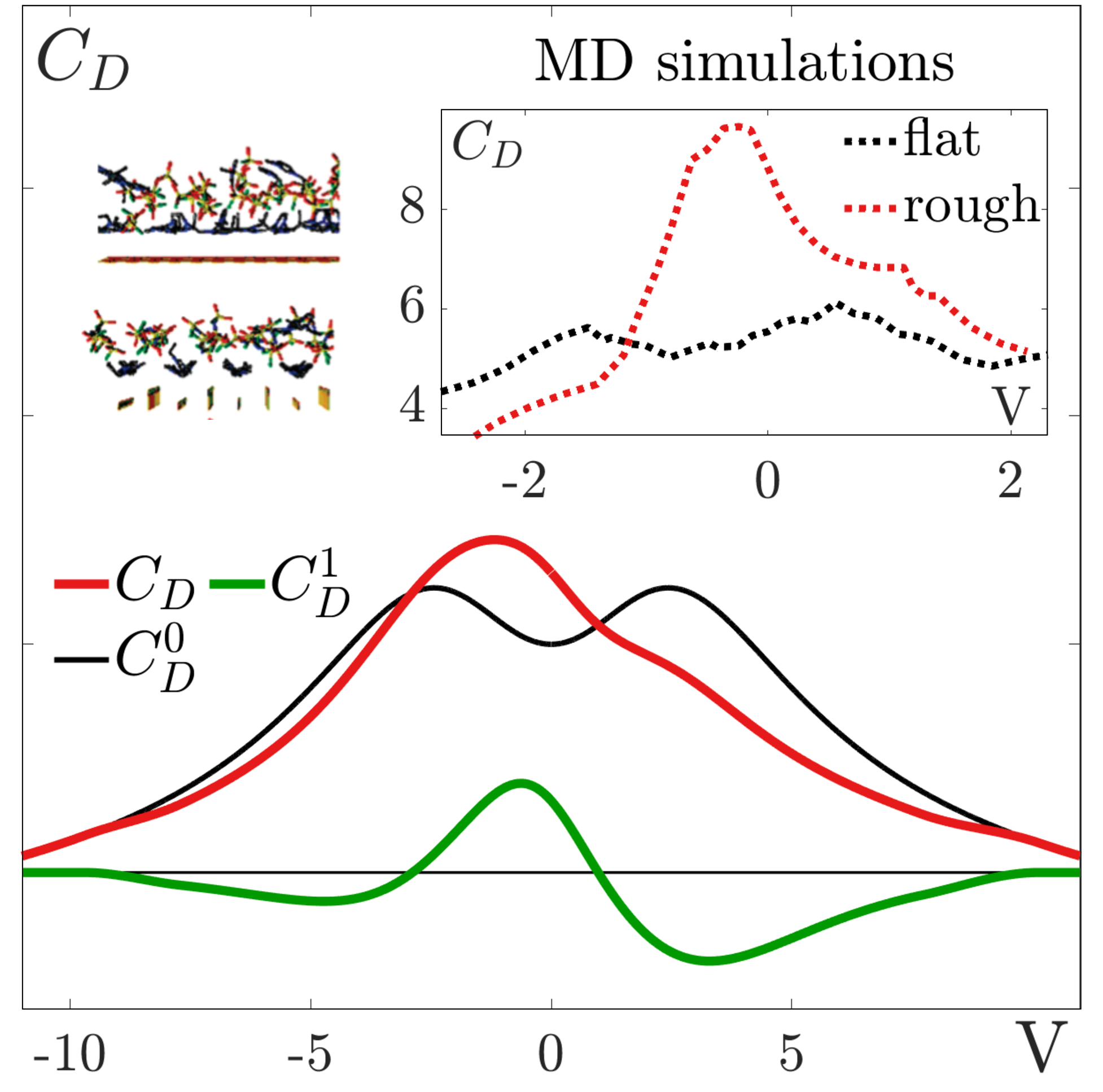}
    \caption{DC profile transition from camel like to bell like due to roughness perturbation. Inset shows results of MD simulation \cite{vatamanu2011influence}.}
    \label{fig:DC_transform}
\end{figure}

\begin{figure*}
    \centering
    \includegraphics[width=1\linewidth]{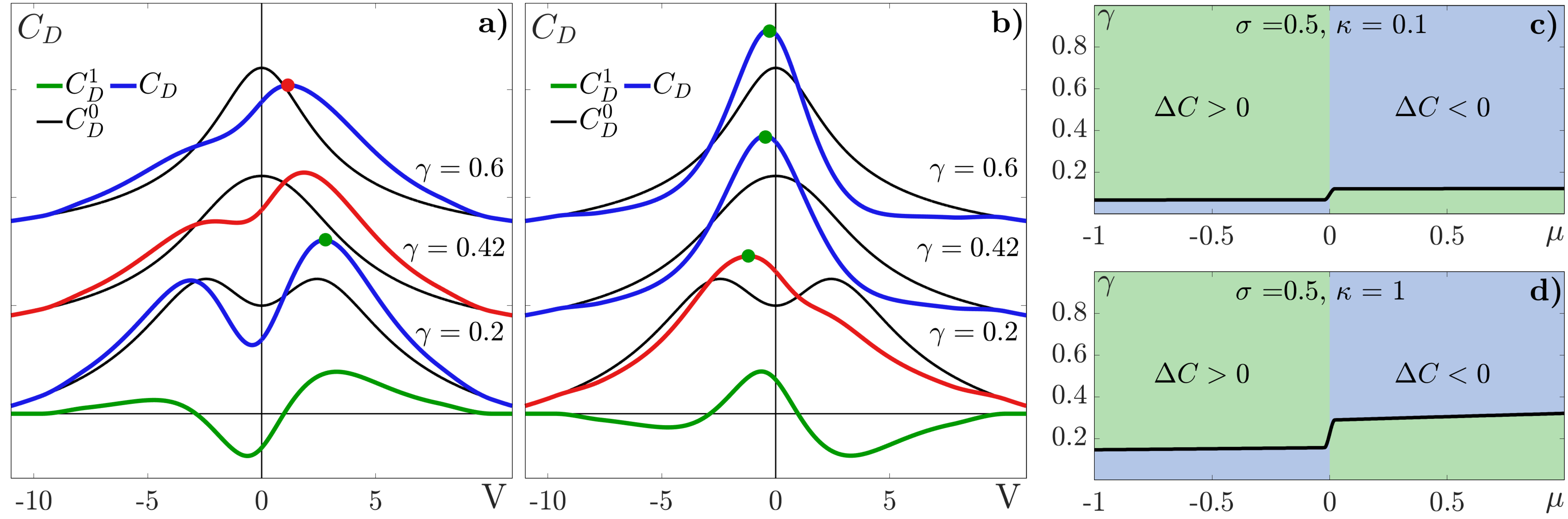}
    \caption{DC profiles a) with positive perturbation b) with negative perturbation at $\mu = 2$ comparing with Kornyshev model for various values of $\gamma$. With a blue and red lines, we plot full DC, black lines correspond to DC from Kornyshev model \cite{kornyshev2007double} without roughness perturbation and, with green line, we illustrate perturbations driven by roughness of electrode surface. We highlight DC with a red line, when perturbation modifies its profile form. Effect of perturbation on DC maximum value c)  at $\sigma = 0.5, \kappa = 0.1$, d) at $\sigma = 0.5, \kappa = 1$. Green area corresponds to DC enhancement and blue one to DC maxima decrease.}
    \label{fig:optimal_design}
\end{figure*}

\subsubsection{Comparison with molecular simulation by Vatamanu et al. \cite{vatamanu2011influence}}

Fig.\ref{fig:DC_transform} shows the DC form transition from camel to bell due to roughness perturbation in the RS model. Here, we consider IL with bigger cations and $\gamma = \gamma_1 + \gamma_2 = 0.2$. We take $L_D$ = 2 nm, $\sigma$ = 2 nm, and $\mu = 2$. This result is compared to MD simulations \cite{vatamanu2011influence}, where DC of EMIM$^+$ FSI$^-$ ionic liquid near flat and rough (prismatic) graphite surface was obtained. Considered IL cation is bigger than anion (EMIM$^+$ ion size is 0.76 nm, FSI$^-$ is about 0.6 nm according to \cite{tiruye2016performance}). The scale of surface roughness is about 0.143 nm, which is smaller than ions sizes. Results of simulations show that camel shape DC profile near the flat electrode turns to a bell shape when the surface became corrugated. The authors explain it by a strong influence of surface topology on IL structure in EDL and modification in intermolecular potential energy between IL ions and the electrode surface. Thus, our model results qualitatively coincide with the results of the MD simulation. 

In Fig.\ref{fig:optimal_design}, with red lines, we highlight cases when the DC profile changes form due to roughness induced perturbation from bell shaped to camel (in Fig.\ref{fig:optimal_design}(a) at $\gamma = 0.42$) and from camel shaped to bell shaped one (in Fig.\ref{fig:optimal_design}(b) at $\gamma = 0.2$). So, we also predict DC form modification from bell to camel. Actually, as perturbation has amplitude $\Delta/\sigma$, DC form transition can also take place with an increasing difference of ion sizes as in MD simulations \cite{hu2014comparative}. Besides, it was also observed in our results for DC dependence on temperature, shown in Fig.\ref{fig:DC_with_T}.

Our analytical model for RS confirms that rough electrode surface can modify DC profile form. We obtain the transition of the DC profile from camel to bell for IL with bigger cations and from bell to camel for IL with bigger anions. DC form transition can be induced with a variety of parameters from IL density to ion size asymmetry, temperature, or surface roughness.

\subsection{Insights to optimal EDLC design}

Next, we investigate conditions required for DC enhancement and select parameters of IL and surface roughness that allow to achieve optimal DC properties. The strongest feature of EDLC is higher power density. As higher DC value as more charge can store EDLC. Hence, it is very essential to understand which properties of IL and electrode microstructure give maximum energy storage.

We apply perturbation from the RS model to DC at different $\gamma$, $\sigma$, and $\kappa$ parameters. So we vary ions density, surface roughness, and Debye length. In Figs.\ref{fig:optimal_design}(a-b), we provide DC profiles for the following cases: (a) positive perturbation in the case of bigger anions with $\mu > 0$ and (b) negative perturbation in the case of bigger cations with $\mu < 0$. We accept $\gamma = \gamma_1 + \gamma_2$. For these calculations, we take $L_D$ = 2 nm, $\sigma$ = 2 nm, and $\mu = \pm 2$. Since perturbation changes are small for different $\gamma$, we plot perturbation only for $\gamma = 0.2$. As can be seen in Fig.\ref{fig:optimal_design}(a), perturbation (green line) can both enhance and decline DC maximum, depending on IL compacity parameter $\gamma$. If $\gamma$ value reaches a particular value of $\gamma^*$, then perturbation will only decline DC maxima value. In Fig.\ref{fig:optimal_design}(b), we obtain only enhancement of DC maximum at considered values of $\gamma$. Therefore, it seems that there are more opportunities in system parameters to enhance DC for IL with bigger cations. In Figs.\ref{fig:optimal_design}(c-d), we illustrate at which value of $\gamma$ perturbation starts to enhance DC maxima. We find that the compacity parameter $\gamma$, at which enhancement switches to diminishing, is usually smaller for a negative value of $\mu$ than for a positive. And for the negative one, we need $\gamma > \gamma^*$ to enhance DC maxima, but for the positive, it should be lower $\gamma < \gamma^*$. We also find that values of $\gamma^*$ rely on the roughness parameter and Debye length. Results of our model predict that the value of $\gamma^*$, at which the effect of perturbation is changed, become bigger for smaller Debye lengths (that is demonstrated in Figs.\ref{fig:optimal_design}(c-d)) and for larger roughness parameter.

We can conclude that for the optimal design of EDLC properties, one should first account IL ion sizes. If IL has bigger cations than for DC enhancement, IL density should be low and provide a smaller realization of inverse Debye length by high temperature and consider smaller roughness. In the case of IL with bigger anions to enhance DC ones should make IL denser, while the inverse Debye length should be higher as surface roughness. Nonetheless, one should create roughness as small till it becomes the same range as a molecular scale to create a stronger modification of DC.

\section{Conclusion}

Molecular scale effects including ion size asymmetry and electrode surface roughness significantly contribute to ions separation in EDL. In this paper, we have proposed novel analytical models that provide PZC and DC solutions for EDL in contact with the rough electrode surface. The models are distinguished by the reason for ions separation: due to the difference of ion sizes and due to electrode surface roughness. Within modest 1D problems, we represent plenty of DC phenomena, observed in the recent numerical and experimental studies, given below:

\begin{enumerate}
    \item Our analytical solution is the first that predicts the formation of the third peak in a DC profile. To obtain it, we include the structural transition of IL ions within the model by modification of the penetration depth parameter to the function of electrode potential. This result reproduces DC profile patterns obtained in the MD study, where ion reorientation causes a complex DC profile.
    \item We obtain the DC profile shape transition from bell to camel and backward due to the roughness of an electrode surface. Roughness creates perturbation that modifies a DC profile. Our model shows that bell-to-camel transition occurs for IL with bigger anions and camel-to-bell for bigger cations. Such behavior was observed in the numerical and experimental studies before.
    \item DC results show monotonic behavior with temperature and ion size ratio for the ISS model and nonmonotonic for the RS model, which could be one of the possible explanations for the disagreement of the previous works.
    \item The obtained PZC results show different signs of PZC values opposite to a bigger ion of IL. For small Debye lengths, PZC approaches the largest in magnitude constant value. Besides, the PZC value grows with temperature and ion size asymmetry. Moreover, the moment when PZC reaches the limit can be controlled with electrode surface roughness. 
    \item We also analyze the parameters of IL and surface roughness required to improve EDLC properties. Remarkably, ion size relation is crucial, and it determines optimal properties. To enhance DC for IL with bigger anions it requires smaller Debye lengths and bigger roughness, while for IL with bigger cations inverse conditions are favorable.  

\end{enumerate}
   
Current work confirms the importance of electrode surface roughness effects on the capacitance characteristics of EDL. The results of the proposed analytical solutions reveal some problems of treating roughness influence on DC behavior. Further, it is of theoretical interest to include as a reference solution the modified Kornyshev model \cite{goodwin2017mean} that account short range ion correlations. We hope that the provided analytical models become useful for the design of optimal EDLC tools.
\nocite{*}
\bibliography{supercapacitors_roughness}
\appendix

\section{ISS model DC}\label{sec:ISS_DC}

To describe DC in ion size separation model, we need to solve the following systems on electrostatic potential:

\begin{equation}\label{eq:iss_1area_problem}
\left\{
\begin{array}{rcl}
\frac{\partial^2 u^I}{\partial x^2} &=& -\frac{1}{2}\frac{e^{-u^I}}{1 + \sum_i \gamma_i \left[ e^{-Z_i u^I} - 1\right]}\\
    u^I|_{x=0} &=& V
\end{array}
\right.
\end{equation}

\begin{equation}\label{eq:iss_2area_problem}
\left\{
\begin{array}{rcl}
\frac{\partial^2 u^{II}}{\partial x^2} &=& \frac{\sinh(u^{II})}{1 + 2\gamma \sinh^2(u^{II}/2)}\\
    u^{II}|_{\frac{\Delta}{L_D}} &=& u^{I}|_{\frac{\Delta}{L_D}} \\
    u_x^{II}|_{\frac{\Delta}{L_D}} &=& u_x^{I}|_{\frac{\Delta}{L_D}}
\end{array}
\right.
\end{equation}
which are formulated for different areas: first area $x \in \left[0, \Delta/L_D \right)$ and second area $x \in \left[\Delta/L_D, +\infty \right)$.

Let's solve the system in the first area. We suggest small $u$ due to high temperatures of the system and low particle density. Then, we can follow classical Debye\,--\,Huckel theory and use linearization of  exp:
\begin{equation}\label{eq:iss_pot1}
    \frac{\partial^2 u^I}{\partial x^2} = -\frac{1}{2} + \frac{1}{2}\left( 1 - \gamma_1 + \gamma_2\right) u^I(x)
\end{equation} 
The solution of Eq.\ref{eq:iss_pot1} has the form:
\begin{equation}\label{eq:iss_u1_solution_1} 
    u^I(x) = \frac{1}{2\lambda^2} + C_1 e^{\lambda x} + C_2 e^{- \lambda x}
\end{equation}
Here $\lambda^2 = (1-\gamma_1 + \gamma_2)/2$. Boundary condition gives: $\frac{1}{2\lambda^2} + C_1 + C_2 = V$.

Moving to the second area, we do not need to solve this system. It will be enough to find the first derivative of $u(x)$ in this area, that was obtained in \cite{kornyshev2007double}:
\begin{equation}
    \frac{\partial u^{II}}{\partial x} = \mp \sqrt{\frac{2}{\gamma}} \sqrt{\log\left(1+2\gamma \sinh^2\left(\frac{u^{II}}{2}\right)\right)}
\end{equation} 
Here, minus for positive charged electrode, and plus --- for negative one. To achieve analytical formulation, Kornyshev model is simplified to Gouy--Chapman approximation for small $u$:
\begin{equation}
    \frac{\partial u^{II}}{\partial x} = - 2sign(u^{II}) \|\sinh \left(\frac{u^{II}}{2}\right)\|  = - 2 \sinh \left(\frac{u^{II}}{2}\right)
\end{equation}

The final system to solve is:
$$\left\{
    \begin{aligned}\label{eq:iss_system1}
    1/2\lambda^2 + C_1 &+ C_2 = V\\
    \left.\frac{\partial u^I}{\partial x} \right|_{x=\frac{\Delta}{L_D}} &= \left.\frac{\partial u^{II}}{\partial x} \right|_{x=\frac{\Delta}{L_D}}
    \end{aligned} \right.$$
Accounting that $u^{II}|_{\frac{\Delta}{L_D}} = u^{I}|_{\frac{\Delta}{L_D}}$, $\Delta/L_D = \alpha$ and substituting Eq.\ref{eq:iss_u1_solution_1}, one can obtain:
$$\left\{
    \begin{aligned}\label{eq:iss_system_2}
     1/2\lambda^2 + C_1 &+ C_2 = V\\
    \lambda C_1 e^{\lambda \alpha} - \lambda C_2 e^{-\lambda \alpha}  &= \\
    -2 \sinh ( &0.5\left[1/2\lambda^2 + C_1 e^{\lambda \alpha} + C_2 e^{-\lambda \alpha}\right])
    \end{aligned} \right.$$
 Expressing $C_2$ from the first equation of system and substituting in the second equation, we obtain following:
\begin{equation}
\begin{aligned}
     C_1 \cosh(\lambda \alpha) - \frac{1}{2} (V-\frac{1}{2\lambda^2})e^{-\lambda \alpha} &= \\
    -\frac{1}{\lambda}\sinh \Bigl(\frac{1}{4 \lambda^2}+\frac{1}{2}(V-\frac{1}{2\lambda^2})e^{-\lambda \alpha}&+C_1 \sinh(\lambda \alpha)\Bigr) \\
\end{aligned}
\end{equation}
\begin{figure*}
\centering
\subfigure[]{
\includegraphics[width=0.36\linewidth]{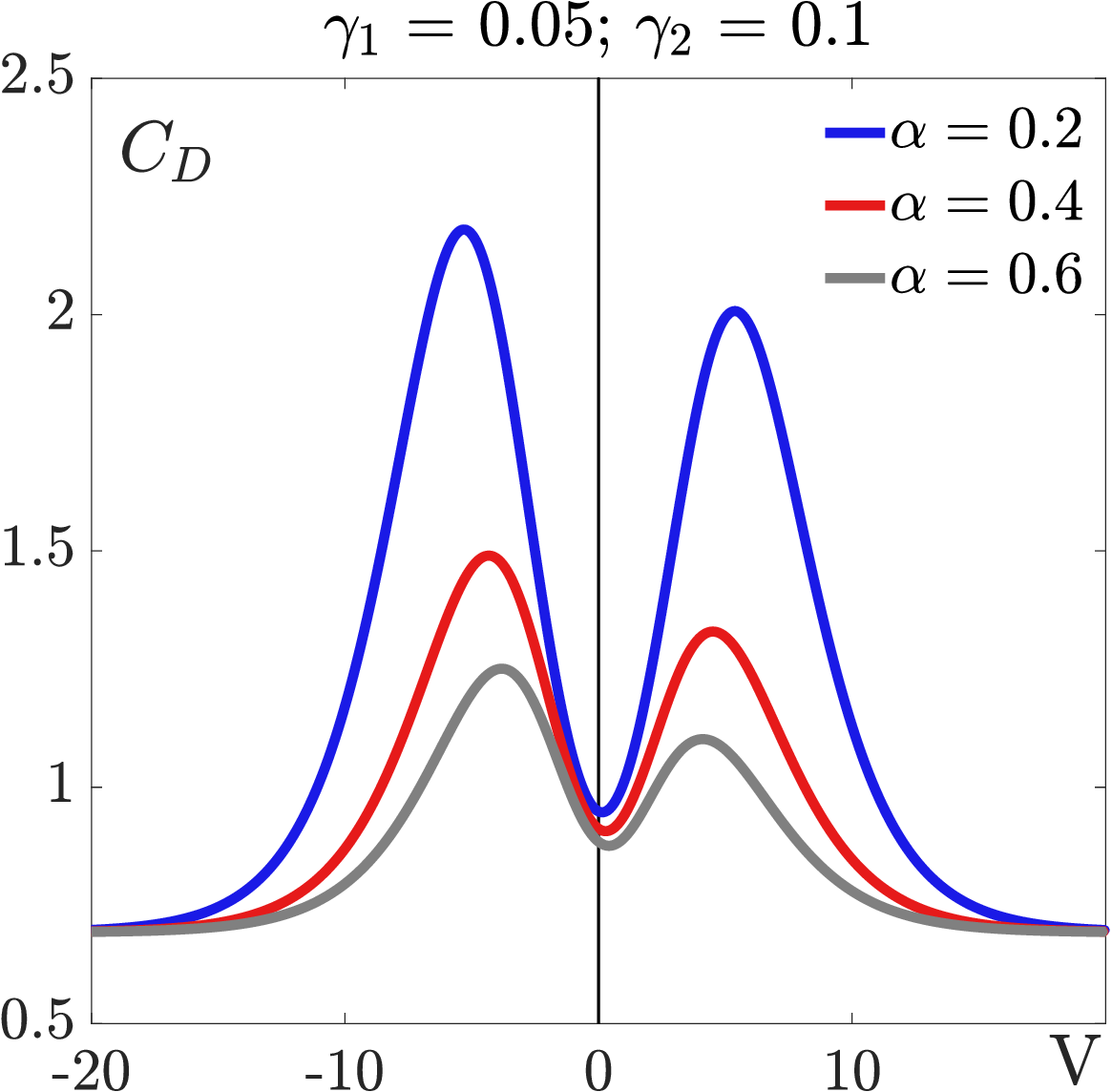} 
\label{fig:DC_first_case} } 
\hspace{0cm}
\subfigure[]{
\includegraphics[width=0.378\linewidth]{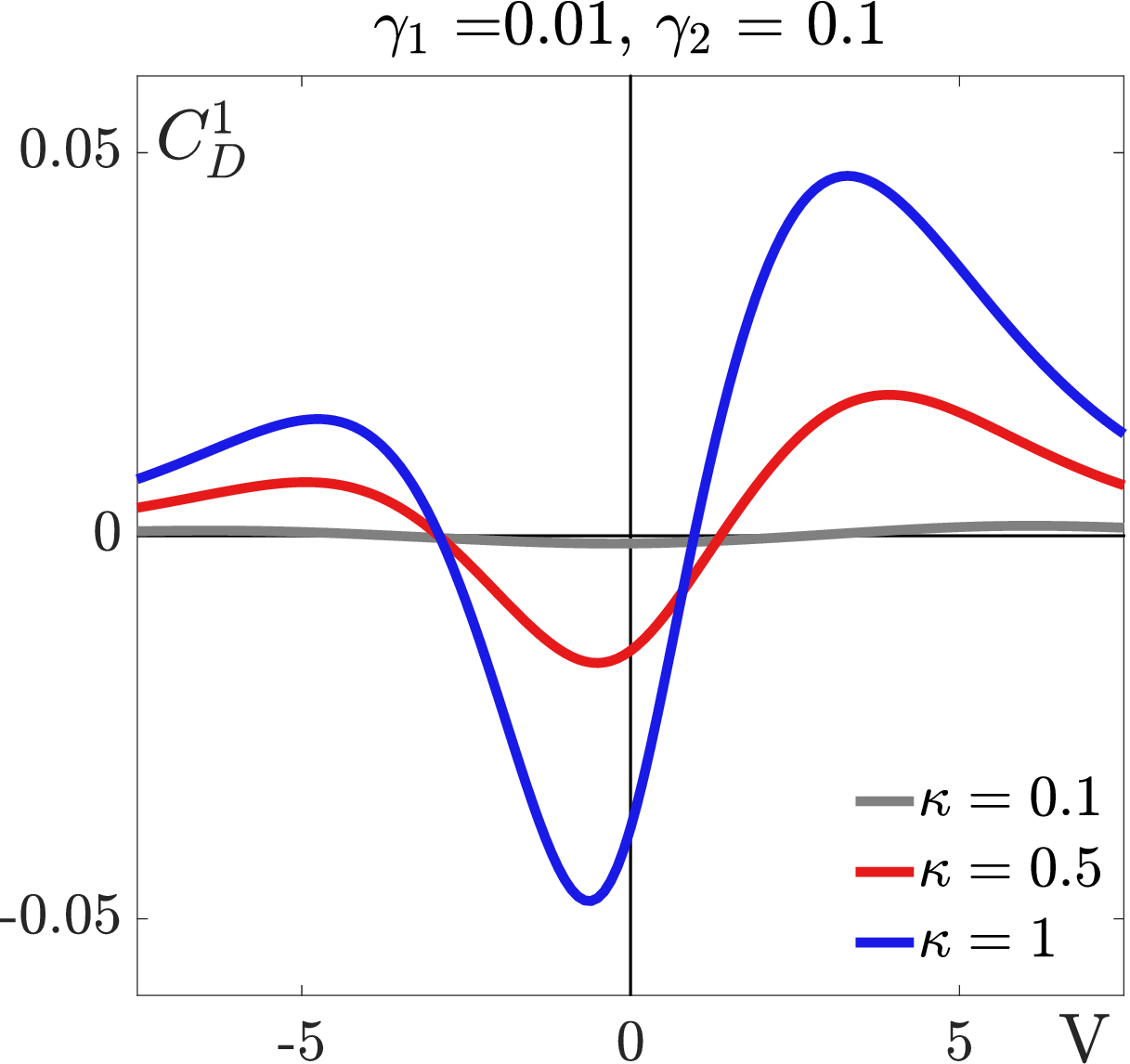}
\label{fig:DC_second_case} }
\caption {(a)~--- DC-potential curves with different $\alpha = \Delta/L_D$ for ISS model ($\sigma \ll \Delta$), (b)~--- DC perturbation vs.potential with different $\kappa$ for RS model ($\sigma \gg \Delta$).}
\end{figure*}
Here, we apply a suggestion that $\Delta \ll L_D$, because $\Delta \sim$~\AA \ and $L_D \sim$ 10 nm.
\begin{equation}
    C_1  - \frac{1}{2} \Bigl(V-\frac{1}{2\lambda^2}\Bigr) =
    -\frac{1}{\lambda}\sinh \Bigl(\frac{V}{2} + \alpha \lambda C_1\Bigr)
\end{equation} 
Expanding $\sinh$ of sum, one can obtain:
\begin{equation}
\begin{aligned}
    C_1  - \frac{1}{2} \Bigl(V-\frac{1}{2\lambda^2}\Bigr) = 
    -&\frac{1}{\lambda}\left( \sinh\left(\frac{V}{2}\right) \cosh(\alpha \lambda C_1) \right.\\ &+ \left.\sinh (\alpha \lambda C_1) \cosh \left(\frac{V}{2}\right) \right)
\end{aligned}
\end{equation} 
For $\Delta \ll L_D$, it reduces to:
\begin{equation}
    C_1  - \frac{1}{2} \Bigl(V-\frac{1}{2\lambda^2}\Bigr) =-\frac{1}{\lambda}\left(\sinh\left(\frac{V}{2}\right) +\alpha \lambda C_1 \cosh \left(\frac{V}{2}\right)\right)
\end{equation} 
Here, parameter $C_1$ can be expressed:
\begin{equation}\label{eq:c_1}
    C_1 = \frac{\frac{1}{2}(V-\frac{1}{2\lambda^2})-\frac{1}{\lambda}\sinh(V/2)}{1+\alpha \cosh(V/2)}
\end{equation} 
Then, substituting $C_2 = V - C_1 - 1/2\lambda^2$ into Eq.\ref{eq:iss_u1_solution_1}, we obtain expression for the potential:
\begin{equation}\label{eq:iss_el_pot}
\begin{aligned}
u(x) &= 1/2\lambda^2 + C_1 e^{ \lambda x} + (V-C_1-1/2\lambda^2) e^{- \lambda x}\\
&= 1/2\lambda^2 + (V-1/2\lambda^2) e^{- \lambda x} + 2C_1\sinh(\lambda x)
\end{aligned}
\end{equation}

According to Gauss law $Q \sim -\left.\frac{\partial u}{\partial x} \right|_{x=0}$, then:
\begin{equation}\label{eq:iss_Q1}
    Q \sim \lambda (V-1/2\lambda^2) - 2 \lambda C_1= \lambda (V-1/2\lambda^2-2C_1)
\end{equation}
Substituting Eq.\ref{eq:c_1} into Eq.\ref{eq:iss_Q1}, cumulative charge takes the form:
\begin{equation}\label{eq:iss_Q}
    Q \sim \frac{\alpha\lambda(V-1/2\lambda^2)\cosh(V/2) + 2\sinh(V/2)}{1 + \alpha\cosh(V/2)}
\end{equation}

To derive $C_D$, we need to calculate the following:
\begin{equation}
\label{eq:iss_dc}
    C_D=\frac{\partial Q}{\partial E}=
    \left.-\frac{\epsilon}{4\pi L_D}\frac{\partial}{\partial V}\frac{\partial u(x)}{\partial x}\right|_{x=0}
\end{equation}

\begin{widetext}

Finally, full expression for DC result:

\begin{equation}\label{eq:capacity_1}
    C_D = \frac{\epsilon}{4\pi L_D} \left[ \frac{(1 +\alpha \lambda) \cosh(V/2)}{1+\alpha \cosh(V/2)} + \frac{\alpha \sinh(V/2) ((V-1/2\lambda^2)\lambda/2 - \sinh(V/2))}{ (1+\alpha \cosh(V/2))^2}\right]
\end{equation}

\end{widetext}

DC obtained in the ISS model (see Eq.\ref{eq:capacity_1}) is a function of 3 parameters: $C_D = f(\alpha,\lambda,V)$. The DC--potential profiles are shown for various $\alpha$ values and $\gamma_1 = 0.05, \gamma_2 = 0.1$ in Fig.\ref{fig:DC_first_case}. DC values are normed at a factor $\frac{\epsilon}{4\pi L_D}$. We obtain a camel like profiles with a higher peak for negative potential.  If we consider IL with bigger cations, the higher peak will be at positive potential, that corresponds with the results from \cite{lockett2008differential, aslyamov2021electrolyte, fedorov2008ionic}. Besides, it has a limit at high potential $C_D \to \lambda - 1/\alpha$.
Differential capacitance in Kornyshev model $C_D^{K} \to 0$ at $|V| \to \infty$, as $C_D^K \sim |V|^{-1/2}$. Constant limit of DC was found experimentally in \cite{nanjundiah1997differential}. Unfortunately, this detail was not discussed in these works \cite{nanjundiah1997differential, kornyshev2007double}. We notice that in \cite{nanjundiah1997differential} DC is presented for mercury electrode that means small electrode surface roughness, so it can be represented with this model.

\section{RS model DC}\label{sec:RS_DC}

In this part, we provide details for DC calculation in RS model. Here, we need to solve the following system for potential perturbation:
\begin{equation}\label{eq:RS_problem_perturb}
\left\{
\begin{array}{lll}
\Delta u_1 =& \left.\frac{\partial f}{\partial u}\right|_{u_0} u_1 &- \frac{1}{2} \tilde g(x) e^{u_0}\\
    \left.u_1 \right|_{x=0} &=0&  \\
    \left.u_1 \right|_{x \to \infty} &=0& 
\end{array}
\right.
\end{equation}

Accounting that:
\begin{equation}
    \left.\frac{\partial f}{\partial u}\right|_{u_0} = \gamma_1 + \gamma_2 + \cosh(u_0)(1 + \gamma_1 + \gamma_2) = \lambda^2 \nonumber
\end{equation}

The first equation in system \ref{eq:RS_problem_perturb} becomes screened Poisson equation, and it can be solved using Green function.
\begin{align}
    [\Delta-\lambda^2] u_1 = &- \frac{1}{2} \tilde g(x) e^{u_0}\label{eq:SPE}\\
    [\Delta-\lambda^2] G_{\infty}(x,x_0)  & = \delta(x - x_0)\label{eq:Gauss_SPE}
\end{align}
Here, $G_{\infty}(x,x_0)$ is "free space" Green's function:
\begin{align}
    G_{\infty}(x,x_0) &= -\frac{1}{2\lambda} e^{-\lambda |x-x_0|}\\
   \lim_{x \to \pm \infty} &G_{\infty}(x,x_0) = 0
\end{align}

To treat boundary conditions of system \ref{eq:RS_problem_perturb}, we need to use "method of images" to find Green function that will help us obtain the solution.
\begin{equation}
\begin{aligned}
     G(x,x_0) = G_{\infty}(x,x_0) - G_{\infty}(-x,x_0) \\
     =-\frac{1}{2\lambda} \left(e^{-\lambda |x-x_0|} - e^{-\lambda |x+x_0|}\right)
\end{aligned}
\end{equation}
Then, potential perturbation $u_1$ takes the form:
\begin{equation}
    u_1(x) = \frac{1}{4\lambda} \int_0^{\infty} \tilde g(x_0) e^{u_0(x_0)}\left[e^{-\lambda |x-x_0|} - e^{-\lambda |x+x_0|}\right] d x_0
\end{equation}

Next, let's evaluate charge perturbation:
\begin{equation} \label{eq:RS_charge_petrub}
Q_1 \sim - \mu \left.\frac{\partial u_1}{\partial x}\right|_{x=0} =
 -\frac{\Delta}{2 \sigma} \int_0^{\infty} \tilde g(x_0) e^{u_0(x_0)}e^{-\lambda x_0} d x_0
\end{equation}
We substitute $u_0$ according to Kornyshev solution. To move further analytically, we can approximate in assumption of small $V \ll 1$:
\begin{equation}
    e^{u_0(x_0)} \approx 1+4\tanh(V/4)e^{-x_0}
\end{equation}
Substituting it in previous equation, one can calculate:
\begin{equation}
\begin{aligned}
    \left.\frac{\partial u_1}{\partial x}\right|_{x=0} = &\frac{1}{2} \int_0^{\infty} \tilde g(x_0) e^{-\lambda x_0} d x_0 + \\ &2\tanh(V/4)\int_0^{\infty} \tilde g(x_0) e^{-(\lambda+1) x_0} d x_0 
\end{aligned}
\end{equation}
Let's replace $\tilde g(x_0)$ by it definition and change variable on $t = x_0 L_D/\sqrt{2} \sigma$, then:
\begin{equation}
\begin{aligned}
    \left.\frac{\partial u_1}{\partial x}\right|_{x=0} = &\frac{1}{2\sqrt{\pi}} \frac{\sigma}{L_D} \left[\int_0^{\infty} e^{- t^2 -\lambda \sqrt{2} \frac{\sigma}{L_D}t}\right. d t + \\ &\left.4\tanh\left(\frac{V}{4}\right)\int_0^{\infty} e^{- t^2 -(\lambda+1) \sqrt{2} \frac{\sigma}{L_D}t} d t \right] \\
    = \frac{1}{2\sqrt{\pi}} &\frac{\sigma}{L_D} \left[ F\left(\sqrt{2}\lambda \frac{\sigma}{L_D}\right)\right. + \\ &\left.4\tanh\left(\frac{V}{4}\right) F\left(\sqrt{2}(\lambda + 1) \frac{\sigma}{L_D}\right) \right]
\end{aligned}
\end{equation}
where $F(s) = \mathcal L[e^{-t^2}]$ --- Laplace transform of Gaussian curve.
\begin{equation}
    F(s) = \int_0^{\infty} e^{-t^2-st} dt = \frac{\sqrt{\pi}}{2}e^{\frac{s^2}{4}} \erfc \left(\frac{s}{2}\right)
\end{equation}

Then, 
\begin{equation} \label{eq:RS_charge_perturb}
\begin{aligned}
    Q_1 \sim - &\mu \left.\frac{\partial u_1}{\partial x}\right|_{x=0} = -\frac{\Delta}{4L_D} \left[e^{\frac{\lambda^2 \sigma^2}{2 L_D^2}} \erfc\left(\frac{\sqrt{2}\lambda}{2} \frac{\sigma}{L_D}\right)  + \right. \\&\left.4\tanh\left(\frac{V}{4}\right)e^{\frac{(\lambda + 1)^2 \sigma^2}{2 L_D^2}} \erfc\left(\frac{\sqrt{2}(\lambda+1)}{2} \frac{\sigma}{L_D}\right) \right]
\end{aligned}
\end{equation}

\textit{Case A. Small Debye lengths.} If $\sigma \gg L_D$, we can use asymptotic expansion of the complementary error function for large real argument: 
\begin{equation}
    \erfc(x) = \frac{e^{-x^2}}{x\sqrt{\pi}} \sum_{n=0}^{\infty} (-1)^n \frac{(2n-1)!!}{(2x^2)^n} \approx \frac{e^{-x^2}}{x\sqrt{\pi}} 
\end{equation}
Finally, charge perturbation is defined as:
\begin{equation}
    Q_1 \sim - \frac{1}{2\sqrt{2\pi}} \frac{\Delta}{\sigma} \left[ \frac{1}{\lambda} + 4 \tanh \left(\frac{V}{4}\right)\frac{1}{1+\lambda}\right]
\end{equation}

\textit{Case B. Large Debye lengths.} If $\sigma \ll L_D$, we apply expansion of the complementary error function for small positive real argument: 
\begin{equation}
    \erfc(x) \approx e^{-x^2}\Bigl(1-\frac{2\sqrt{\pi x^2}}{\pi}\Bigr) \approx e^{-x^2}
\end{equation}
In this case, charge perturbation $Q_1$ takes the form:
\begin{equation}
\begin{aligned}
    Q_1 \sim  -\frac{\Delta}{4L_D} & \left[\left(1 - \frac{\sqrt{2\lambda^2}}{\sqrt{\pi}} \frac{\sigma}{L_D}\right)  + \right. \\&\left.4\tanh\left(\frac{V}{4}\right) \left(1 - \frac{\sqrt{2(\lambda+1)^2}}{\sqrt{\pi}} \frac{\sigma}{L_D}\right) \right] \\
    &\approx -\frac{\Delta}{4L_D} \left[1  + 4\tanh\left(\frac{V}{4}\right) \right]
\end{aligned}
\end{equation}

Now, we move to DC formulation. In this case, $C_D = C_D^0 + C_D^1$, where $C_D^0 = \frac{\epsilon}{4\pi L_D} \cosh(V/2)$. The potential derivative of charge perturbation from Eq.\ref{eq:RS_charge_perturb} gives: 

\begin{widetext}

\begin{equation}\label{eq:capacity_2}
\begin{aligned}
      &C_D^1 =  -\frac{\Delta}{4L_D}\frac{\epsilon}{4\pi L_D}\left[\left(\frac{\lambda\sigma^2}{L_D^2}e^{\frac{\lambda^2 \sigma^2}{2 L_D^2}} \erfc\left(\frac{\sqrt{2}\lambda}{2} \frac{\sigma}{L_D}\right) - \sqrt{\frac{2}{\pi}}\frac{\sigma}{L_D}\right) \frac{\partial \lambda}{\partial V} + \right.\\
      4 \tanh \left(\frac{V}{4} \right)\left(\frac{(\lambda+1)\sigma^2}{L_D^2}e^{\frac{(\lambda+1)^2 \sigma^2}{2 L_D^2}} \right.& \left.\erfc\left(\frac{\sqrt{2}(\lambda+1)}{2} \frac{\sigma}{L_D}\right) - \sqrt{\frac{2}{\pi}}\frac{\sigma}{L_D}\right) \frac{\partial \lambda}{\partial V} +
      \frac{1}{\cosh^2(V/4)}  e^{\frac{(\lambda+1)^2 \sigma^2}{2 L_D^2}} \left.\erfc\left(\frac{\sqrt{2}(\lambda+1)}{2} \frac{\sigma}{L_D}\right)\right]
\end{aligned}
\end{equation}
where $\lambda^2 = \gamma_1 + \gamma_2 + \cosh(V)(1 + \gamma_1 + \gamma_2)$ and $\partial \lambda / \partial V = (1 + \gamma_1 + \gamma_2) \sinh(V) / \lambda$.

For small Debye lengths, it reduces to:

\begin{equation}\label{eq:capacity_2_small_L_D}
    C_D^1 = -\frac{1}{2\sqrt{2\pi}}\frac{\Delta}{\sigma}\frac{\epsilon}{4\pi L_D} \left[ \frac{1}{\cosh^2(V/4) (1 + \lambda)} -  \left( \frac{1}{\lambda^2} + 4 \tanh\left(\frac{V}{4}\right) \frac{1}{(1+\lambda)^2}\right)\frac{\partial \lambda}{\partial V}\right]
\end{equation}

For large Debye lengths, it takes the form:
\begin{equation}\label{eq:capacity_2_large_L_D}
     C_D^1 =  -\frac{\Delta}{4L_D} \frac{\epsilon}{4\pi L_D}\left[ -\sqrt{\frac{2}{\pi}}\frac{\sigma}{L_D} \left(1 + 4 \tanh \left(\frac{V}{4} \right)\right)\frac{\partial \lambda}{\partial V} \right. +\left. \left( 1 - \frac{\sqrt{2}(\lambda+1)}{\sqrt{\pi}} \frac{\sigma}{L_D} \right) \frac{1}{\cosh^2(V/4)}\right] \approx - \frac{\Delta}{4L_D} \frac{\epsilon}{4\pi L_D} \frac{1}{\cosh^2(V/4)}
\end{equation}

\end{widetext}

In the RS model, we obtain $C_D^1 = f(\Delta,L_D,\sigma,\lambda,V)$. Fig.\ref{fig:DC_second_case} shows DC perturbation - potential profiles for different inverse Debye lengths. For smaller Debye length, we obtain bigger DC perturbation. Charge perturbation $C_D^1 \to 0$ at large potentials. Perturbation is not symmetric against $V = 0$, its minima is located in area of negative potentials and maxima for positive potentials is bigger than for negative one.

\end{document}